\documentclass[a4paper]{aa_v6.1}
\usepackage{amsmath,amssymb,natbib,graphicx}
\graphicspath{{./Figures/}}
\usepackage{array}

\bibpunct{(}{)}{;}{a}{}{,}

\newcommand{\be}{\begin{equation}}
\newcommand{\ee}{\end{equation}}
\newcommand{\ellipticity}{\epsilon}
\newcommand{\size}{R}
\def\snr{{\mathcal{S}}}

\newcommand{\sigmasys}{\sigma_{\rm sys}}
\newcommand{\bs}{\boldsymbol}

\newcommand{\snreff}{\snr_{\rm eff}}

\def\variance#1{\left<\left|#1\right|^2\right>}
\newcommand{\pun}{P1}
\def\ssqobj#1{{\size^2_{\rm #1}}}

\def\variance#1{\left<\left|#1\right|^2\right>}

\begin{document}

\title{Optimal PSF modeling for weak lensing :\\
complexity and sparsity
}

\author{
S. Paulin-Henriksson\inst{1}, A. Refregier\inst{1}, A. Amara\inst{2,3}
}

\institute{
 Service d'Astrophysique, CEA Saclay, Batiment 709, 91191 Gif--sur--Yvette Cedex, France
\and
Department of Physics, ETH Z\"urich, Wolfgang-Pauli-Strasse 16,  CH-8093 Z\"urich, 
Switzerland.
\and
Department of Physics and Center for Theoretical and Computational Physics, University of Hong Kong, Pok Fu Lam Road, Hong Kong
}

\titlerunning{
Optimal PSF modeling for weak lensing\ldots
}

\authorrunning{
Paulin-Henriksson et al.
}

\date{Received ---; accepted December 23, 2009}

\abstract
{
We address the issue of controling the systematic errors in shape measurements for weak gravitational lensing.
}
{
We make a step to quantify the impact of systematic errors in modeling the point spread function (PSF) of observations, on the determination of cosmological parameters from cosmic shear.}
{
We explore the impact of PSF fitting errors on cosmic shear measurements using the concepts of complexity and sparsity.  Complexity, introduced in a previous paper, characterizes the number of degrees of freedom of the PSF. 
For instance, fitting an underlying PSF with a model of low complexity produces small statistical errors on the model parameters, although these parameters could be affected by large biases.  Alternatively, fitting a large number of parameters (i.e. a high complexity) tends to reduce biases at the expense of increasing the statistical errors. 
We attempt to find a trade-off between scatters and biases by studying the mean squared error of a PSF model. 
We also characterize the model sparsity, which describes how efficiently the model is able to represent the underlying PSF using a limited number of free parameters. 
We present the general case and give an illustration for a realistic example of a PSF modeled by a shapelet basis set.
}
{
We derive a relation between the complexity and the sparsity of the PSF model, the signal-to-noise ratio of stars and the systematic errors in the cosmological parameters. 
By insisting that the systematic errors are below the statistical uncertainties, we derive a relation between the number of stars required to calibrate the PSF and the sparsity of the PSF model.
We discuss the impact of our results for current and future cosmic shear surveys. 
In the typical case where the sparsity can be represented by a power-law function of the complexity, 
we demonstrate that current ground-based surveys can calibrate the PSF with few stars, while future surveys will require hard constraints on the sparsity in order to calibrate the PSF with 50 stars.
}
{
}


\maketitle


\section{Introduction}
\label{sec:intro}

Studying spatial correlations between galaxy shapes induced by gravitational lensing of the large scale structure (`Cosmic Shear') is a powerful probe of dark energy and dark matter. 
A number of current and planned surveys are dedicated to cosmic shear, 
such as: the Canada-France-Hawaii-Telescope Legacy Survey\footnote{\textsf{http://www.cfht.hawaii.edu/Science/CFHLS/}} (CFHTLS), the KIlo Degree Survey and the VISTA Kilo-Degree Infrared Galaxy Survey\footnote{\textsf{http://www.eso.org/sci/observing/policies/\\ \hspace*{1.5cm}PublicSurveys/sciencePublicSurveys.html}} (KIDS/VIKING), the Dark Energy Survey\footnote{\textsf{http://www.darkenergysurvey.org}} (DES), the Panoramic Survey Telescope \& Rapid Response System\footnote{\textsf{http://pan-starrs.ifa.hawaii.edu}} (Pan-STARRS), the SuperNovae Acceleration Probe\footnote{\textsf{http://snap.lbl.gov}} (SNAP), the Large Synoptic Survey Telescope\footnote{\textsf{http://www.lsst.org}} (LSST) and the Dark UNiverse Explorer\footnote{\mbox{\textsf{http://www.dune-mission.net}} and \mbox{\textsf{http://www.esa.int/esaCP/index.html}}} (DUNE/Euclid).

The most efficient way of improving the statistical precision of cosmic shear analyses is by enlarging the surveys. As long as the median redshift is sufficiently high ($z\gtrsim 0.7$), 
\cite{2007MNRAS.381.1018A} demonstrate that, 
it is more advantageous to perform the cosmic shear surveys in fields as wide as possible, rather than deep. This in order to minimize the error bars in cosmological parameters. 
To date, the largest data optimized for cosmic shear is the Wide field of the 
CFHTLS, 
which covers $50\,$deg$^2$ \citep{2008A&A...479....9F} and will eventually reach $170\,$deg$^2$.  
Another analysis has also been published that combines 4 surveys that, together, cover an area of $100\,$deg$^2$ \citep{2007MNRAS.381..702B}. 
In a few years, the KIDS/VIKING survey will cover $1500\,$deg$^2$. Eventually, projects currently being planned, such as DUNE/Euclid, planned for 2017, will be able to perform cosmic shear measurements over the entire observable extragalactic sky ($\sim 20,000\,$deg$^2$).

To let cosmic shear surveys reaching their full potential, it is necessary to ensure that systematic errors are sub-dominant relative to statistical uncertainties. 
In particular, a tight control on all the effects associated with shape measurements is required. 
To illustrate the difficulty in compiling accurate shear measurements, we begin with an overview of  the `forward process' that 
illustrates how the original image of a galaxy is distorted in forming the final image that we measure. 
In the forward process, a galaxy image is: (i) sheared by gravitational lensing; (ii) convolved with a point spread function (PSF) originating in a number of sources (e.g. instruments and atmosphere);  (iii) pixelated at the detector; and finally (iv) affected by noise.

Cosmic shear analyses involve the reverse process: we begin with the final image and move backwards from step (iv) to (i) in recovering the original lensing effect.  A detailed and illustrated description of the forward and inverse processes is given in the GREAT08 Challenge Handbook \citep{2008arXiv0802.1214B}. The GREAT08 Challenge aims to 
provide a wide range of expertise into gravitational lensing by presenting the relevant issues in a clear way so as to make it accessible to a broad range of disciplines, including the machine-learning and signal-processing communities. 
Other similar challenges have also been performed within the weak lensing community, as part of the STEP collaboration \citep{2006MNRAS.368.1323H,2007MNRAS.376...13M,spacestep}, which focused mainly on understanding the systematic errors at play in current shear measurement methods. 
These challenges focus mainly on reducing the errors originating in the shape measurement method. However, even with a perfect method there are 
fundamental limits due the statistical potential of a data set.

In \cite{2008A&A...484...67P} (\pun{} hereafter), we investigated the link between systematic errors in the power spectrum and uncertainties in the PSF correction phase. 
The framework is the following. Since the PSF of an instrument varies on all scales, the PSF needs to be measured using the stars that surround the lensed galaxy. 
Each star provides an image of the PSF that is pixelated and noisy, 
which means that to reach a given accuracy in the knowledge of the PSF, a number of stars is required. 
We estimate the number of stars $N_*$ required to calibrate the PSF to a given accuracy, according to the stellar signal-to-noise ratio (SNR), the minimum galaxy size, the complexity of the PSF and the tolerated variance in the systematics $\sigma^2_{\rm sys}$. 
On the other hand, \cite{2007arXiv0710.5171A} estimated the upper limit to $\sigma^2_{\rm sys}$ 
when estimating cosmological parameters. By combining both papers together, we derive the minimum number of stars required to reach a given accuracy. 
For instance, analyses completed to date, that allow us to constrain $\sigma_8\,\Omega_{\rm M}$ with an accuracy of 0.05, require $\sigma^2_{\rm sys}$ lower than a few $10^{-6}$ and the PSF to be calibrated by using 5 stars; while for future ambitious surveys that will allow us to constrain $w_0$ and $w_a$ to an accuracy of 0.02 and 0.1, respectively, $\sigma^2_{\rm sys}$ must be lower than $10^{-7}$, 
which requires at least 50 stars 
(for stars with signal-to-noise ratio of 500 and a PSF described by a few degrees of freedom, as can be typically achieved in space).

In \pun{} we use the same functional form for both the underlying PSF as well as the model used to fit it. 
This means that the PSF model is able to describe perfectly the underlying PSF. 
The errors in the fit due to noise causes a scatter of the fitted parameters around the truth. For instance, if the model is an orthogonal basis set, then the fitted parameters follow a Gaussian distribution around the truth. 
In this paper, we extend this investigation by studying the impact of fitting a PSF with a model that has a different form. 
This addresses the case in which the underlying PSF (unknown in practice) is estimated by fitting the parameters of an arbitrary model. 
This can lead to both a scatter in the fitting parameters and an offset in the average value relative to the true value, 
i.e. a bias in the fitting parameters. 
We can therefore model a given PSF using either a complex model of small biases but large scatters, or a simpler model that would lead to smaller scatters but larger biases. 
To quantify these effects, we revisit the concept of complexity proposed in \pun{} and introduce the concept of sparsity.

This paper is organised as follow: first, in Sect. \ref{sec:complexityandsparsity}, we discuss  the concepts of complexity and sparsity, which are the key concepts of this paper; 
Sect.$\,$\ref{sec:systematics} presents our notation; 
Sect.$\,$\ref{sec:optpsfmodel} presents the definition of optimal complexity, illustrates our formalism with a PSF example and uses the sparsity as a tool for optimizing the complexity; 
Sect.$\,$\ref{sec:reqnumstars} derives the minimum number of stars required to calibrate the PSF, extending results of \pun ;
 and finally, sect.$\,$\ref{sec:conclusion} summarizes our conclusions.


\section{Complexity and sparsity}
\label{sec:complexityandsparsity}

In \pun , we introduced the concept of complexity; we demonstrated that a few complexity factors characterize the amount of information that needs to be collected about the PSF. 
This is summarized and revisited in Sect.$\,$\ref{sec:complexity}. 
In Sect.$\,$\ref{sec:introsparsity}, we introduce the concept of sparsity, which measures the ability of a PSF model to represent the underlying PSF with a small number of free parameters. 
This allows us to explore how an optimal PSF model can be constructed to minimize $\sigma_{\rm sys}$.


\subsection{Complexity}
\label{sec:complexity}

In \pun , we define the complexity factors of the PSF, which represent the number of degrees of freedom (DoF) that are estimated from stars (in the limit of infinite resolution, i.e. infinitely small pixels): the higher the number of DoF, the larger the complexity factors. 
Each PSF shape parameter is associated with a complexity factor that is related to the rms of its estimator. 
In the simple formalism where we consider unweighted quadrupole moments, the PSF is characterized by only two complexity factors $\psi_\ellipticity$ and $\psi_{\size^2}$ associated with the 2 component PSF ellipticity $\bs{\ellipticity}_{\rm PSF}$ and the square PSF rms radius $\size^2_{\rm PSF}$ respectively (as defined in \pun ). 
For a given star, one has:
\be
\label{eq:1}
\left(
\begin{array}{cc}
\psi_{\size^2}\\
\psi_{\ellipticity}
\end{array}
\right)
\equiv \mathcal{S}_*\,
\left(
\begin{array}{cc}
\sigma[\size^2_{\rm PSF}] / \size^2_{\rm PSF}\\
\sigma[\ellipticity_{\rm PSF}]
\end{array}
\right) 
\ee
where $\mathcal{S}_*$ is the photometric SNR of the star, $\sigma[\size^2_{\rm PSF}]$ is the rms error of the PSF size estimator and $\sigma[\ellipticity_{\rm PSF}]$ is the rms error of the PSF ellipticity estimator. 
As in \pun , we assume that the small ellipticity regime holds (i.e. $|\bs{\ellipticity}_{\rm PSF}|\lesssim 0.1$), implying that the measure of $\bs{\ellipticity}_{\rm PSF}$ is isotropic and the 2 components of the ellipticity have the same rms uncertainty, 
i.e. $\sigma[\ellipticity_{{\rm PSF},i}]\equiv\sigma[\ellipticity_{\rm PSF}]$.

If the PSF can be considered constant over several stars, or for particular representations of the PSF (for example  with shapelet basis sets in the small ellipticity regime, see \pun ), $\psi_{\size^2}$ and $\psi_{\ellipticity}$ are spatially constant and Eq. \ref{eq:1} can be extended to a set of several stars. For a combination of several stars, 
$\mathcal{S}_*$ becomes $\sqrt{n_*}\mathcal{S}_{\rm eff}$:
\be
\label{eq:sigell2sigsize2}
\left(
\begin{array}{cc}
\psi_{\size^2}\\
\psi_{\ellipticity}
\end{array}
\right)
\equiv \sqrt{n_*}\,\snreff\,
\left(
\begin{array}{cc}
\sigma[\size^2_{\rm PSF}] / \size^2_{\rm PSF}\\
\sigma[\ellipticity_{\rm PSF}]
\end{array}
\right)\,,
\ee
where ${\snreff}$ is the effective stellar SNR and 
$n_*$ is the effective number of stars $k$ used in the PSF calibration, as defined by (see \pun ):
\be
\label{eq:snreffdef}
n_*\,\snreff^2 \equiv \sum_k\snr^2_{k}\,.
\ee

We also show in \pun{} that the polar shapelet basis set, proposed by \cite{2005MNRAS.363..197M}, tested on simulated data in \cite{2007MNRAS.376...13M} and used on real data by \cite{2008MNRAS.385..695B}, is particularly convenient for modeling the PSF. For example, in the small ellipticity regime, $\psi_\ellipticity$ and $\psi_{\size^2}$ depend only on the polar shapelet basis set over which the PSF is decomposed, not on the PSF itself. 
For this reason we use shapelets in this paper when illustrating our discussions by an example. 
Note that our results and conclusions are not restricted to shapelets but remain valid whatever the PSF model. 
For convenience, we choose the shapelet `diamond' option (described in details in \pun ) that imposes a lower limit to the scales described, implying a link between $\psi_\ellipticity$ and $\psi_{\size^2}$:
\begin{eqnarray}
\psi_\ellipticity^2 & = & \psi_{\size^2}^2 - \mathcal{N}\,,
\end{eqnarray}
with $\mathcal{N}$ the highest even integer lower than or equal to the order $n_{\rm max}$ of the basis set. 
We then consider the overall complexity $\Psi$ defined in the following according to $\psi_\ellipticity$, $\psi_{\size^2}$ and the variance in the galaxy ellipticity distribution.


\subsection{Sparsity}
\label{sec:introsparsity}

In this paper, we introduce the concept of ÔsparsityÕ of the PSF model, which describes how efficiently a model can represent the underlying PSF with a limited number of DoF (i.e. with a limited complexity). 
Specifically, the sparsity quantifies how the residuals between the estimated and the underlying PSF decrease as the complexity of the PSF model increases. 
With a high number of DoF, i.e. a high complexity, one might expect small residuals but large scatters ointhe fitted parameters. 
On the other hand, with a small number of DoF, i.e. a low complexity, one might expect large residuals but small scatters in the fitted parameters. 
The sparsity characterizes the slope in this relation and 
thus is an estimate of 
the amount of information that can be contained in a given number of DoF. 
We show how to use sparsity in optimizing the complexity and minimizing $\sigma_{\rm sys}$.

Consider the shape parameters $\size^2_{\rm PSF}$ and $\bs{\ellipticity}_{\rm PSF}$ of the underlying PSF, as defined previously. 
The differences $\delta(\size^2_{\rm PSF})$ and $\bs{\delta\ellipticity}_{\rm PSF}$ between the underlying PSF (`true' index) and its estimation (`est' index) can be written:
\be
\begin{array}{lcl}
\delta(\size^2_{\rm PSF}) & \equiv & \size^2_{\rm est} - \size^2_{\rm true}\,,\\
\bs{\delta\ellipticity}_{\rm PSF} & \equiv & \bs{\ellipticity}_{\rm est} - \bs{\ellipticity}_{\rm true}\,.
\end{array}
\ee
These differences are of two types: 
the statistical scatter relative to the average $\sigma$ and the bias-offset $b$ of the average relative to the true value. 
The mean square errors (MSE) of $\size^2_{\rm PSF}$ and $\bs{\ellipticity}_{\rm PSF}$ are:
\begin{eqnarray}
\label{eq:sigmasizedef}
{\rm MSE}[\size^2_{\rm PSF}] & \equiv & \sigma^2[\size_{\rm PSF}^2] + b^2[\size_{\rm PSF}^2]\;,\\
\label{eq:sigmaelldef}
{\rm MSE}[\ellipticity_{{\rm PSF},i}] & \equiv & \sigma^2[\ellipticity_{{\rm PSF},i}] + b^2[\ellipticity_{{\rm PSF},i}]\;{\rm with}\,i=1,2\,.
\end{eqnarray}
In \pun , we address the zero bias case \mbox{$b[\ellipticity_i]=b[\size^2]=0$}, that is equivalent to consider the PSF model is able to describe the underlying PSF perfectly. 
However, this nulling of the biases is not necessarily the optimal PSF modeling. It can be advantageous to work with a simplistic PSF model that is unable to describe all the PSF features and has some biases but low statistical scatter (see our PSF example in Sect. \ref{sec:psfex} and Fig. \ref{fig:sparsepsf}). 
This paper proposes another approach that consists in optimizing the PSF model in order to minimize $\sigma_{\rm sys}$. We do this by searching for the optimal trade-off between the systematic errors ($b[\ellipticity_1]$, $b[\ellipticity_2]$, and $b[\size^2]$) and the statistical errors ($\sigma[\ellipticity]$, and $\sigma[\size^2]$), 
which is equivalent to searching the optimal complexity $\Psi$ of the PSF model. 
The sparsity allows us to perfom this optimization because it characterizes the decrease in the biases $b$ as $\Psi$ increases.

In the following, we define a `sparsity parameter' $\alpha$ in the particular case where the biases are modeled as a power-law function of the PSF model complexity ($B\propto1/\Psi^\alpha$), and we study the impact of $\alpha$ on the number of stars required to calibrate the PSF. We thus revisit the main result of \pun{} by deriving $N_*$ (the number of stars required to calibrate the PSF) according to $\alpha$ instead of $\Psi$ (the complexity of the underlying PSF). 
Moreover, this new relation is optimized to minimize $\sigma_{\rm sys}$. 

We emphasize that, in this paper, we propose to optimize the complexity of the PSF model within a given the basis set. 
We do not address the issue of choosing the basis set itself. There is no doubt that, to optimize the PSF modeling, it is necessary to select carefully this basis set. 
For instance, generic basis sets such as shapelets, wavelets, or Fourier modes, although they have enormous advantages, are not 
optimal. 
This issue will be addressed in forthcoming works.


\section{PSF calibration for shear measurement}
\label{sec:systematics}

When deconvolving the observed galaxy with the estimated PSF, $\delta(\size^2_{\rm PSF})$ and $\bs{\delta\ellipticity}_{\rm PSF}$ propagate into an 
error $\bs{\delta\ellipticity}_{\rm gal}$ in the estimation of the galaxy ellipticity. We denote $\size_{\rm gal}$ and $\bs{\ellipticity}_{\rm gal}$, the rms radius and the two-component ellipticity of the galaxy. When $\size_{\rm PSF}$, $\bs{\ellipticity}_{\rm PSF}$, $\size_{\rm gal}$, and $\bs{\ellipticity}_{\rm gal}$ are defined using the unweighted moments of the flux, this propagates to (see P1):
\be
\label{eq:errorpropagationpsfintogal}
\bs{\delta\ellipticity}^{\rm sys}_{\rm gal} \simeq 
\left( \bs{\ellipticity}_{\rm gal} - \bs{\ellipticity}_{\rm PSF} \right)
 \frac{\delta\left(\ssqobj{PSF}\right)}{\ssqobj{gal}} - \left(\frac{\size_{\rm PSF}}{\size_{\rm gal}}\right)^2 \bs{\delta\ellipticity}_{\rm PSF}\,.
\ee
The spatial average of $|\bs{\delta\ellipticity}^{\rm sys}_{\rm gal}|^2$ is related to the variance in the systematic errors in the shear measurements $\sigma_{\rm sys}^2$ \citep{2007arXiv0710.5171A}, defined by the integral of the systematics $C^{\rm sys}_\ell$ in the power spectrum:
\begin{eqnarray}
\label{eq:sigmasysdef6}
\sigma_{\rm sys}^2 & \equiv & \frac{1}{2\pi}\int d(\ln\,\ell)\,\ell(\ell+1)\,\left|C^{\rm sys}_\ell\right|\\
\label{eq:sigmasysdef1}
 & \equiv & \left(P^\gamma\right)^{-2} \variance{ \bs{\delta\ellipticity}^{\rm sys}_{\rm gal}}
\end{eqnarray}
with $P^\gamma$ the calibration factor between the gravitational shear and the ellipticity. Its value depends on the distribution of galaxy ellipticities and is typically about $1.84$ \citep{2000ApJ...536...79R}. The brackets $<\,>$ denote a spatial average over the entire field. 
As in \pun , we substitute Eq. \ref{eq:errorpropagationpsfintogal} into Eq. \ref{eq:sigmasysdef1} with the following simplifying assumptions:
\begin{enumerate}
\item The galaxy is not correlated with the PSF.
\item The error on the PSF ellipticity ($\bs{\delta\ellipticity}_{\rm PSF}$) and the PSF ellipticity itself ($\bs{\ellipticity}_{\rm PSF}$) are not correlated. This is warranted by the fact that, in the assumed small ellipticity regime, $\bs{\delta\ellipticity}_{\rm PSF}$ does not have any preferred direction, implying that \mbox{$<\bs{\ellipticity}_{\rm PSF}.\bs{\delta\ellipticity}_{\rm PSF}>=0$}.
\item We neglect correlations between the ellipticity and the inverse squared radius of the galaxy. This is reasonable for the PSF calibration in the small ellipticity regime.
\end{enumerate}
With these simplifications, one can substitute Eq. \ref{eq:errorpropagationpsfintogal} into Eq. \ref{eq:sigmasysdef1} to obtain:
\begin{eqnarray}
\label{eq:sigmasysdef}
\sigma_{\rm sys}^2 & = & \left(P^\gamma\right)^{-2} \left<\left(\frac{\size_{\rm PSF}}{\size_{\rm gal}}\right)^4\right> \times
     \Bigg[ \variance{ \bs{\delta\ellipticity}_{\rm PSF} } +\nonumber\\
 & & \left[ \variance{ \bs{\ellipticity}_{\rm gal} } + \variance{ \bs{\ellipticity}_{\rm PSF} } \right]
     \frac{\variance{ \delta\size^2_{\rm PSF} }}{\left<\size^4_{\rm PSF}\right>}
     \Bigg]\,.
\end{eqnarray}
We develop a more compact expression by adopting the following notation:
\begin{eqnarray}
\label{eq:cdef}
\mathcal{C} & \equiv & \left(P^\gamma\right)^{-2} \left<\left(\frac{\size_{\rm PSF}}{\size_{\rm gal}}\right)^4\right>\,,\\
\label{eq:edef}
\mathcal{E} & \equiv & \variance{ \bs{\ellipticity}_{\rm gal} } + \variance{ \bs{\ellipticity}_{\rm PSF} }\,,
\end{eqnarray}
which leads to:
\be
\label{eq:sigmasysdef4}
\sigma_{\rm sys}^2 = \mathcal{C}\left[\left<\left|\bs{\delta\ellipticity}_{\rm PSF}\right|^2\right> + \frac{\mathcal{E}}{\size^4_{\rm PSF}}\left<\left|\delta\size^2_{\rm PSF}\right|^2\right>\right]\,.
\ee

In \pun , we considered only the scatters $\sigma[\size_{\rm PSF}^2]$ and $\sigma[\ellipticity_{{\rm PSF}}]$ (i.e. in the zero bias case: \mbox{$b[\ellipticity_i]=b[\size^2]=0$}) and we approximate the statistical averages with spatial averages:   
\mbox{$\sigma^2[\size^2_{\rm PSF}]\simeq\left<\left|\delta\size^2_{\rm PSF}\right|^2\right>$} and 
\mbox{$\sigma^2[\ellipticity_{{\rm PSF},i}]\simeq\left<\left|\delta\ellipticity_{{\rm PSF},i}\right|^2\right>$}. 
In this paper, with the introduction of biases, the scatter becomes the MSE:
\be
\label{eq:2}
\begin{array}{lcl}
{\rm MSE}[\size^2_{\rm PSF}] & \simeq & \left<\left|\delta\size^2_{\rm PSF}\right|^2\right>\,,\\
{\rm MSE}[\ellipticity_{{\rm PSF},i}] & \simeq & \left<\left|\delta\ellipticity_{{\rm PSF},i}\right|^2\right>\,.
\end{array}
\ee
This leads to:
\be
\label{eq:sigmasysdef2}
\begin{array}{lcl}
\sigma_{\rm sys}^2 & \simeq & \mathcal{C}\left[b^2[\ellipticity_{{\rm PSF},1}] + \sigma^2[\ellipticity_{{\rm PSF},1}]\right.\\
& & + b^2[\ellipticity_{{\rm PSF},2}] + \sigma^2[\ellipticity_{{\rm PSF},2}]\\
& & + \left.\frac{\mathcal{E}}{\size^4_{\rm PSF}}\left(b^2[\size^2_{\rm PSF}] + \sigma^2[\size^2_{\rm PSF}]\right)\right]
\end{array}\,.
\ee
We can see that $\sigma_{\rm sys}^2$ is proportional to the quadratic sum of 6 terms: 
three bias terms and three statistical ones. 
Collecting terms of similar type using the following notation:
\begin{eqnarray}
\label{eq:tapta}
B & \equiv & b^2[\ellipticity_1^{\rm PSF}] + b^2[\ellipticity_2^{\rm PSF}] + \mathcal{E}\frac{b^2[\size_{\rm PSF}^2]}{\size_{\rm PSF}^4}\,,\\
\Sigma & \equiv & \sigma^2[\ellipticity_1^{\rm PSF}] + \sigma^2[\ellipticity_2^{\rm PSF}] + \mathcal{E}\frac{\sigma^2[\size^2_{\rm PSF}]}{\size_{\rm PSF}^4}\,,
\end{eqnarray}
gives:
\be
\label{eq:sigmasysdef5}
\sigmasys^2 \simeq \mathcal{C}\left[B + \Sigma\right]\,.
\ee
Although only $\Sigma$ depends on the SNR of the stars,
B and $\Sigma$ both depend on the complexity of the modeling. They can be denoted $B(\Psi)$ and $\Sigma(\Psi,\snreff)$. In \pun , we show that it is given by:
\be
\label{eq:totpo}
\Sigma(\Psi,\snreff) = \frac{\Psi^2}{n_*\mathcal{S}^2_{\rm eff}}\,,
\ee
where the overall complexity of the model $\Psi$ is given by the complexities $\psi_\ellipticity$ and $\psi_{\size^2}$ associated with the ellipticity and the squared radius of the model, respectively (see Eq. \ref{eq:sigell2sigsize2}):
\be
\label{eq:bigpsidef}
\Psi^2 = 2\psi_\ellipticity^2 + \mathcal{E}\psi^2_{\size^2}\,.
\ee
Equations \ref{eq:sigmasysdef5} and \ref{eq:totpo} then infer that:
\be
\label{eq:sigmasysdef3}
\sigmasys^2 \simeq \mathcal{C}\left[B(\Psi) + \frac{\Psi^2}{n_*\,\snreff^2}\right]\,.
\ee
From this equation, we can see that increasing the complexity $\Psi$ by adding degrees of freedom in the PSF model can reduce $B$ but also increases the statistical errors. Minimizing $\sigmasys^2$ thus implies the search for the optimal trade-off in the value of $\Psi$.


\section{Optimal PSF model}
\label{sec:optpsfmodel}

In Sect.$\,$\ref{sec:psfex}, we present a PSF example that we use in the remainder of this paper to illustrate our discussion. 
In Sect.$\,$\ref{sec:optpsfmod}, we show the optimal complexity of the PSF model (that minimizes $\sigma_{\rm sys}$) and apply this to the PSF example. 
We then explore this optimization in more details in Sect.$\,$\ref{sec:sparsity} by examining a particular case in which the bias can be described by a power-law function of the complexity.


\subsection{PSF example}
\label{sec:psfex}

\begin{figure*}[!tbp]
\begin{center}
\includegraphics[scale=0.4]{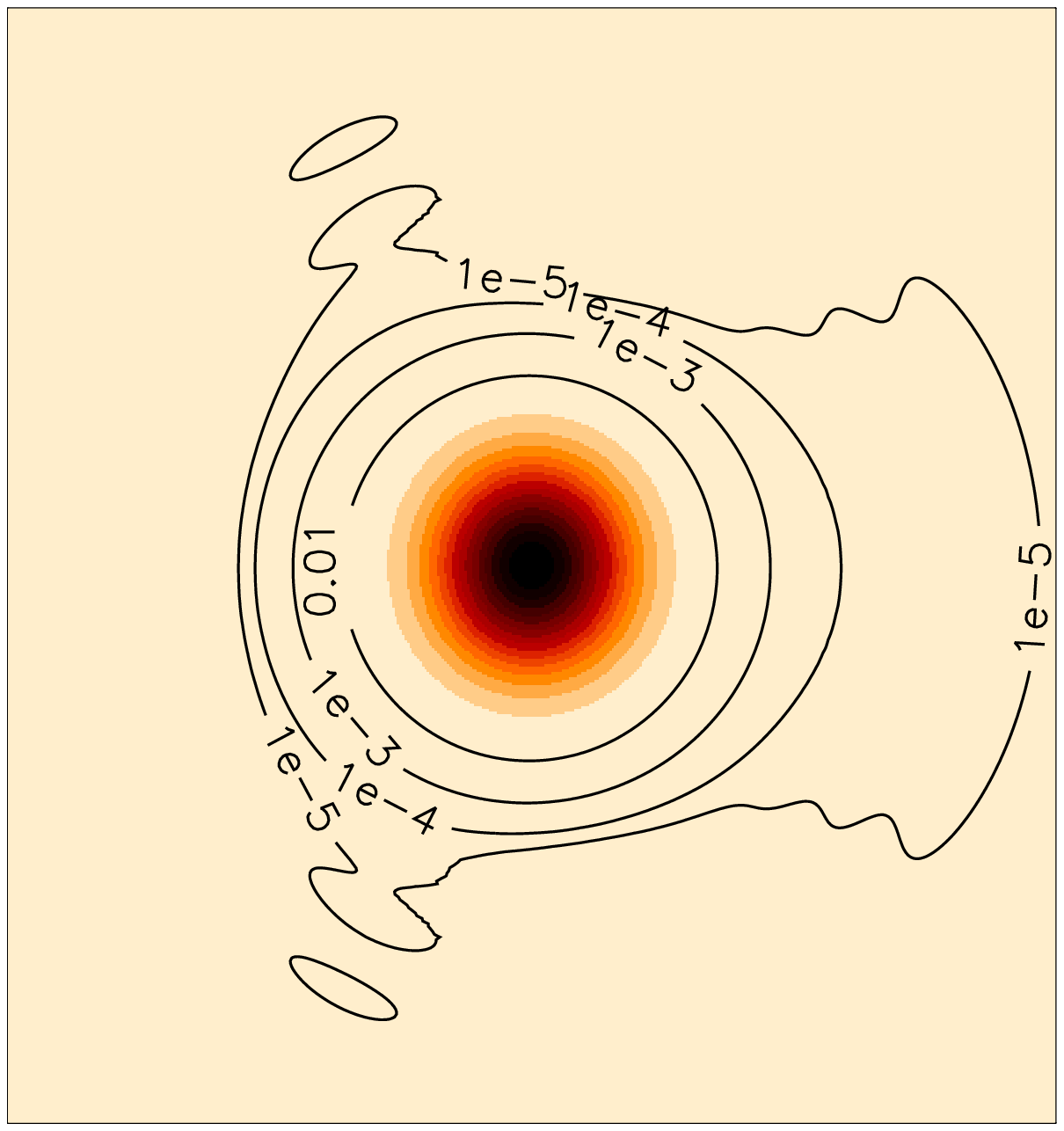}
\end{center}
\vspace*{-6.5cm}
\begin{center}
\hspace*{4cm}original\\
\hspace*{4cm}($\Psi=28.4$)
\end{center}
\vspace*{4.2cm}
\begin{tabular}{c m{2.7cm} m{2.7cm} m{2.7cm} m{2.7cm}}
 & \hspace*{1cm}$\Psi_{\rm fit}=2.6$ & \hspace*{1cm}$\Psi_{\rm fit}=4.3$ & \hspace*{1cm}$\Psi_{\rm fit}=7.8$ & \hspace*{1cm}$\Psi_{\rm fit}=16.4$\\
\begin{tabular}{c}
$\sqrt{n_*}\mathcal{S}_{\rm eff}=\infty$\\
$(\Psi_{\rm opt} = 28.4$)
\end{tabular} & 
\includegraphics[scale=0.25,bbllx=60,bblly=38,bburx=408,bbury=408]{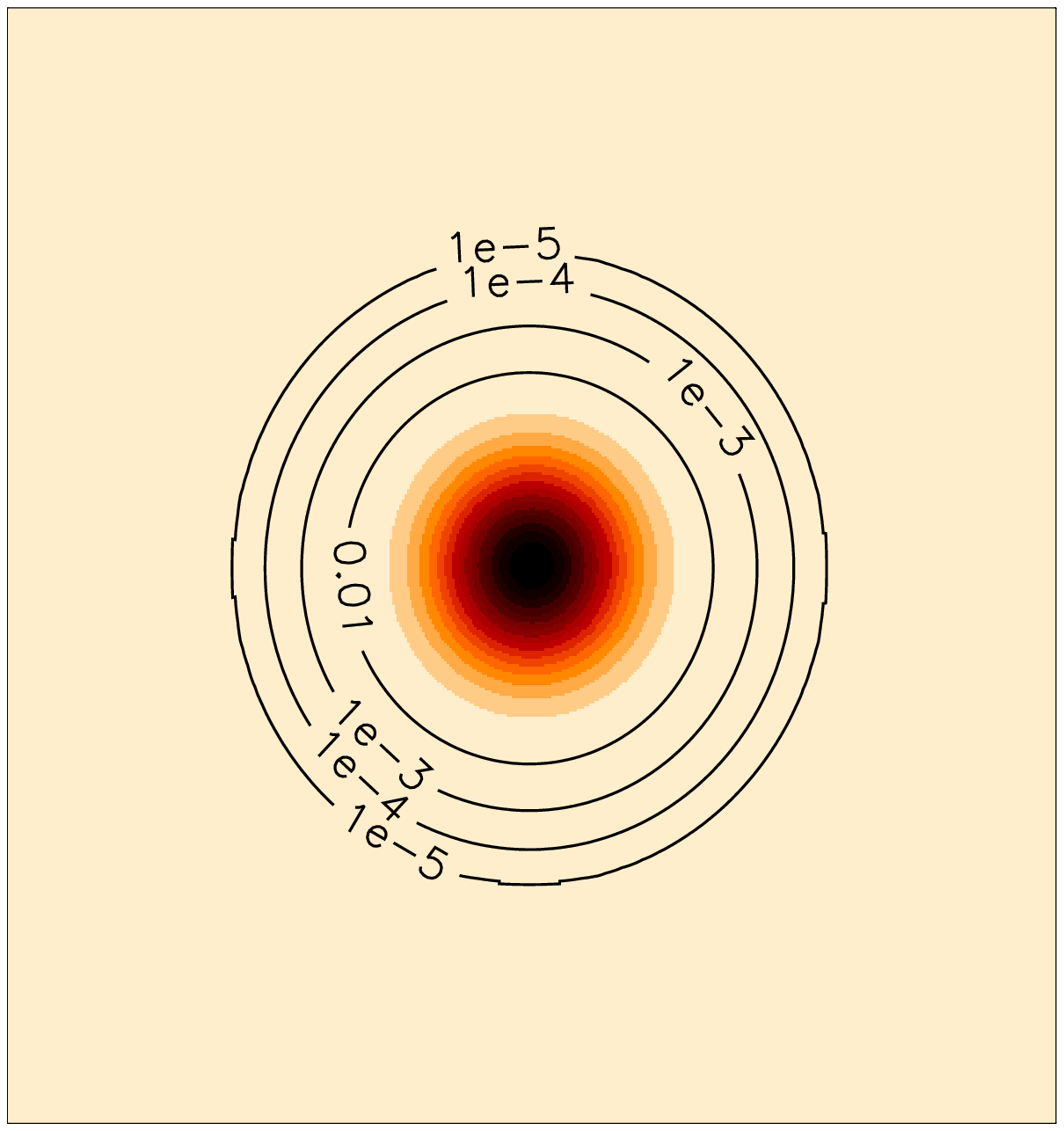} &
\includegraphics[scale=0.25,bbllx=60,bblly=38,bburx=408,bbury=408]{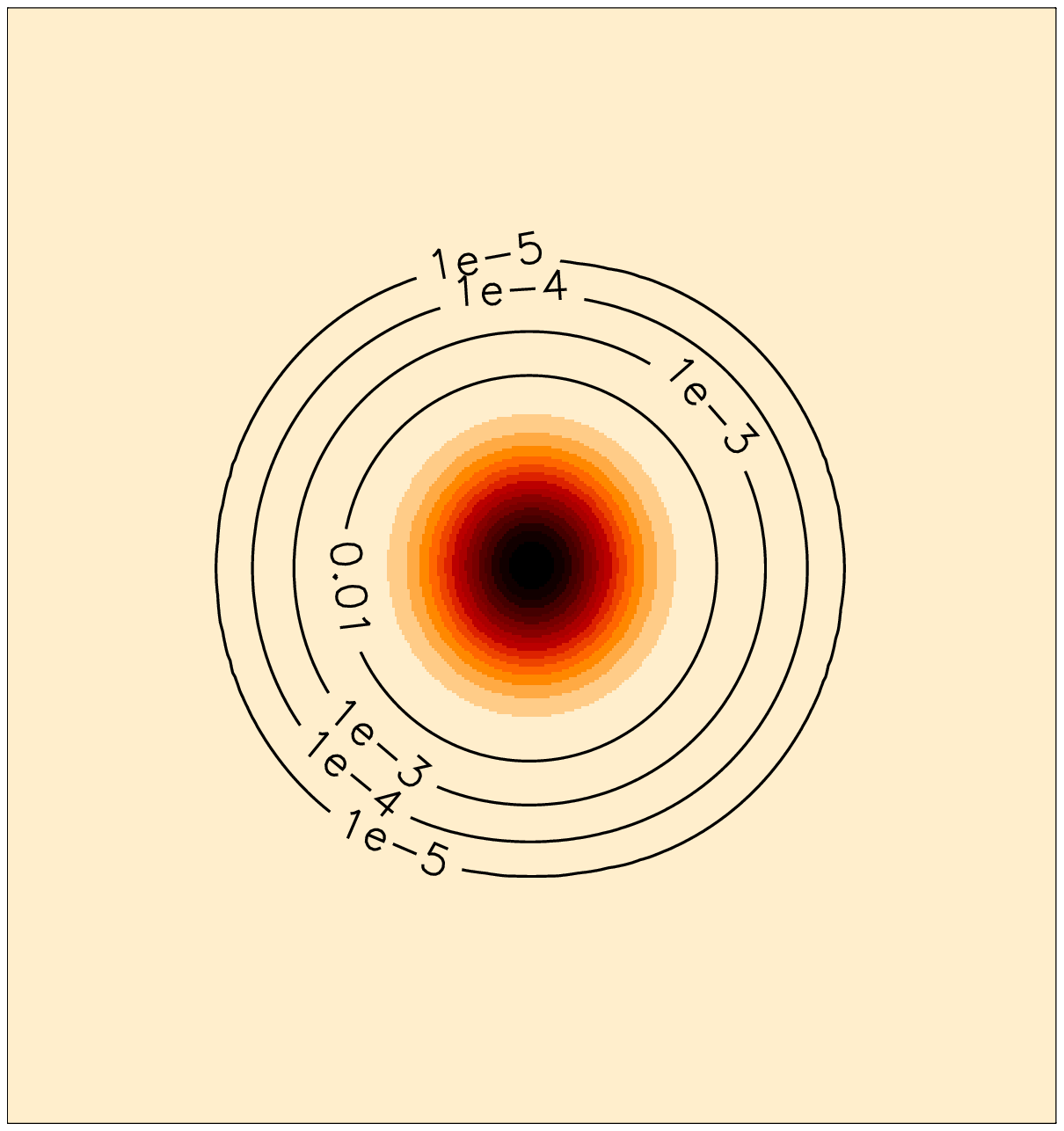} &
\includegraphics[scale=0.25,bbllx=60,bblly=38,bburx=408,bbury=408]{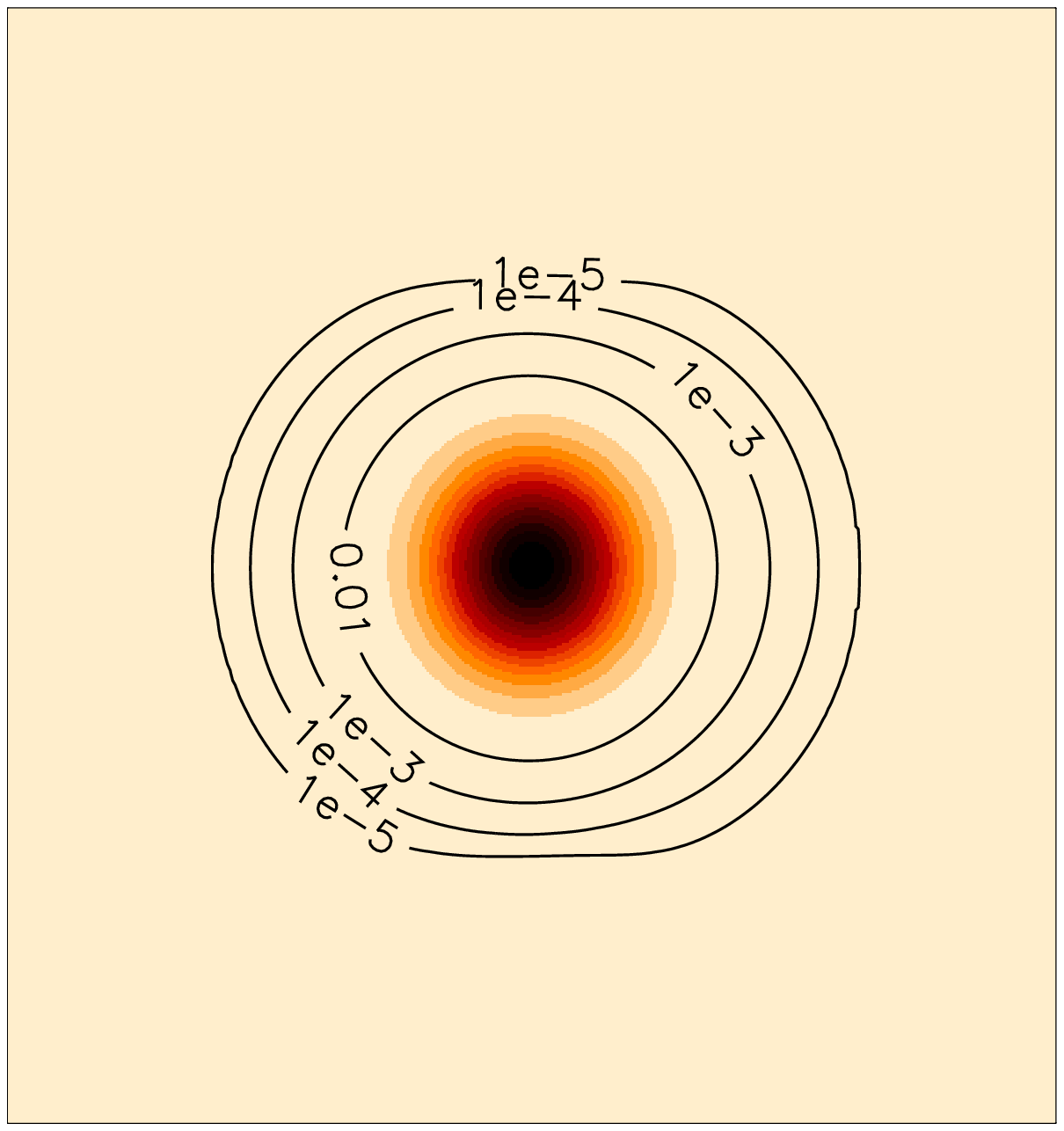} &
\includegraphics[scale=0.25,bbllx=60,bblly=38,bburx=408,bbury=408]{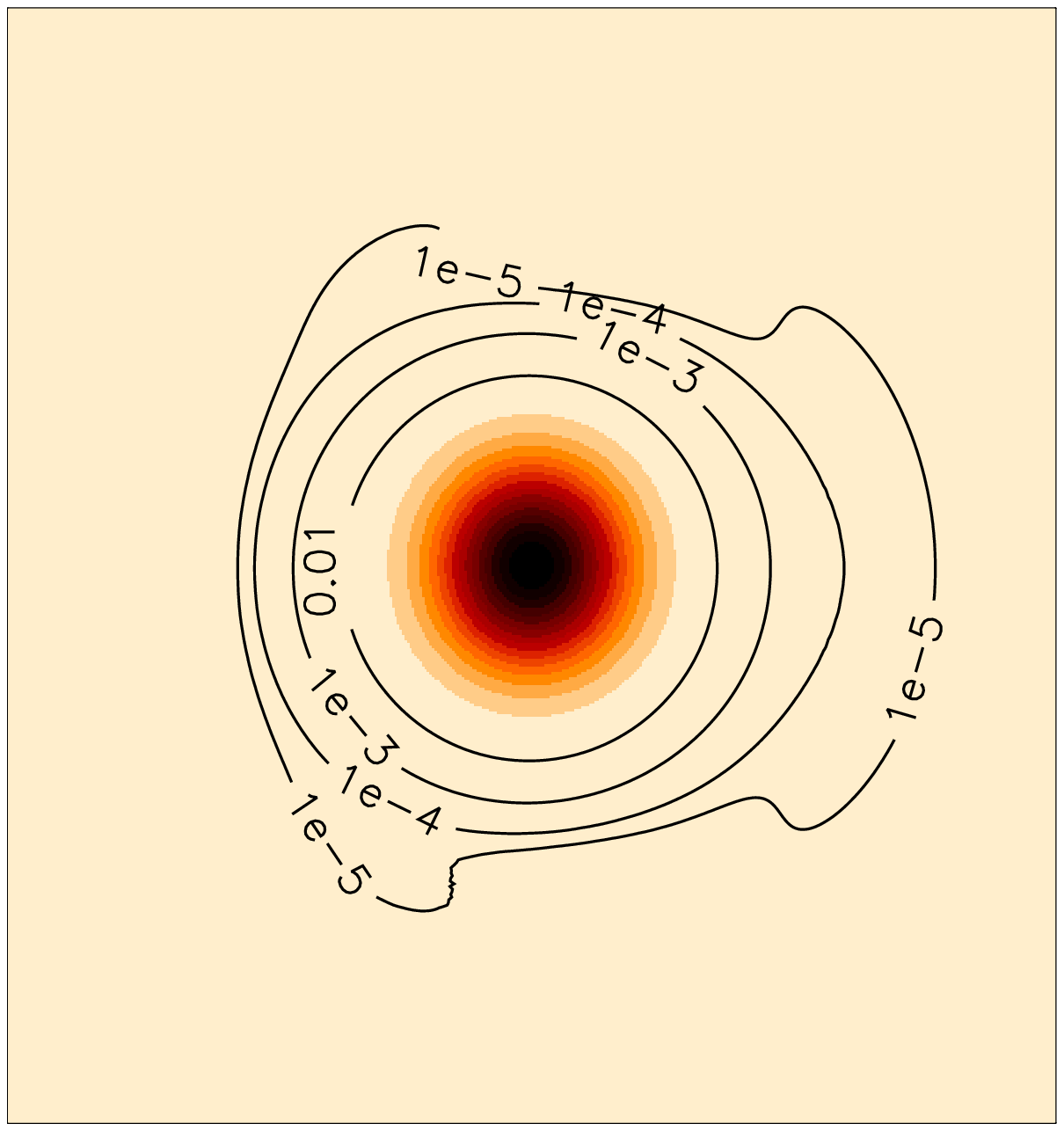}\\
\begin{tabular}{c}
$\sqrt{n_*}\mathcal{S}_{\rm eff}=10^4$\\
$(\Psi_{\rm opt}=7.8$)
\end{tabular} & 
\includegraphics[scale=0.25,bbllx=60,bblly=38,bburx=408,bbury=408]{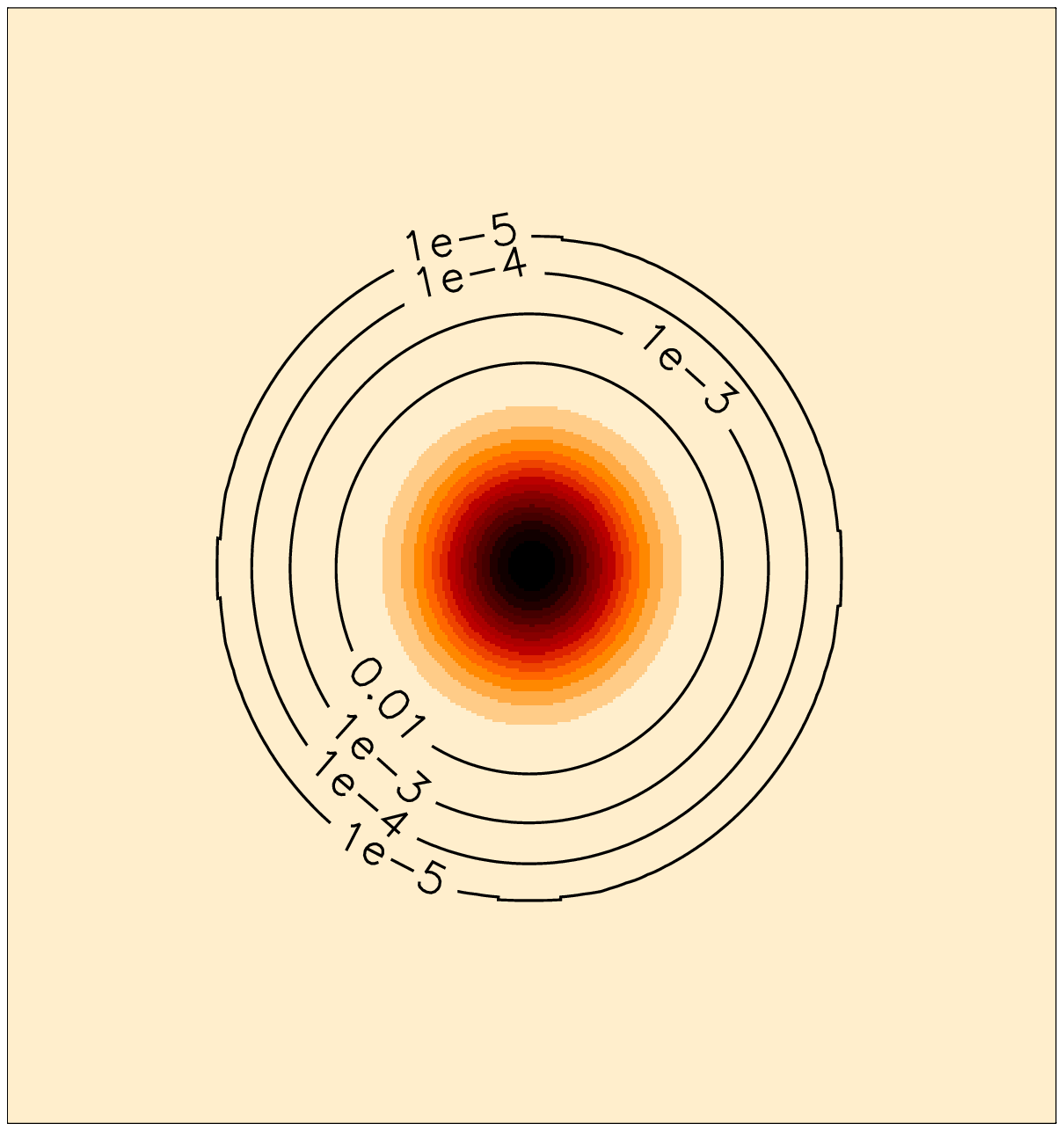} &
\includegraphics[scale=0.25,bbllx=60,bblly=38,bburx=408,bbury=408]{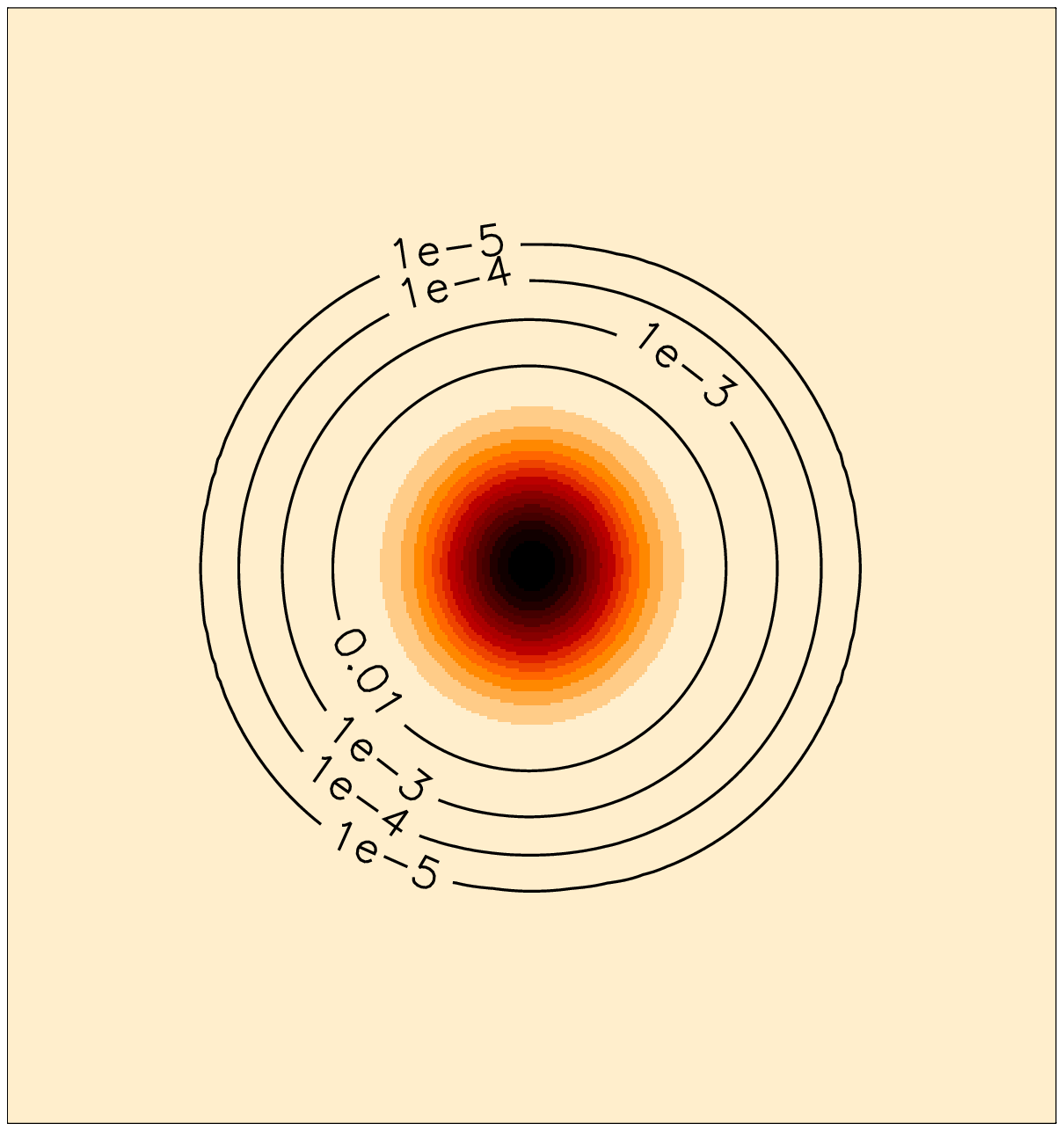} &
\includegraphics[scale=0.25,bbllx=60,bblly=38,bburx=408,bbury=408]{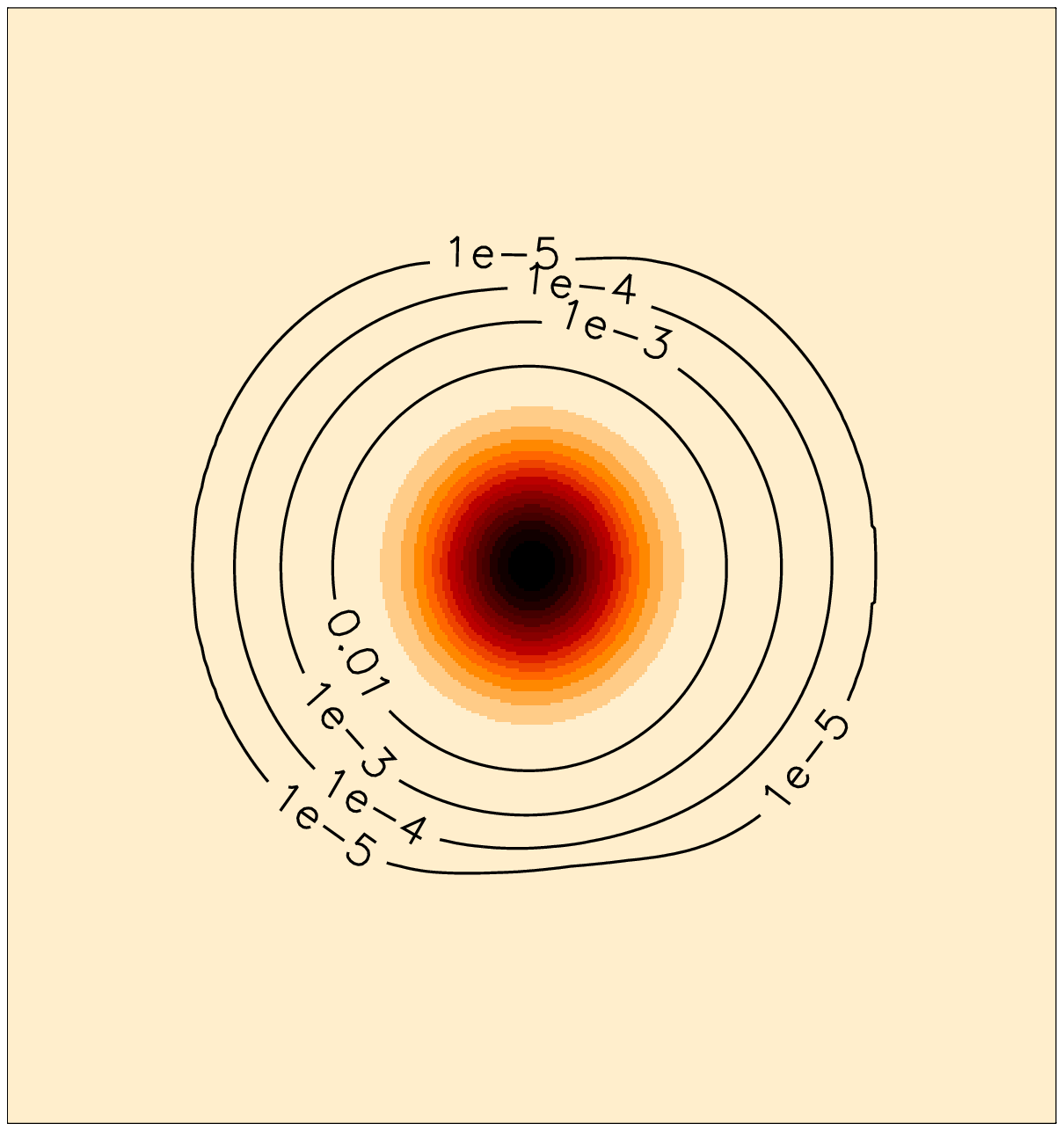} &
\includegraphics[scale=0.25,bbllx=60,bblly=38,bburx=408,bbury=408]{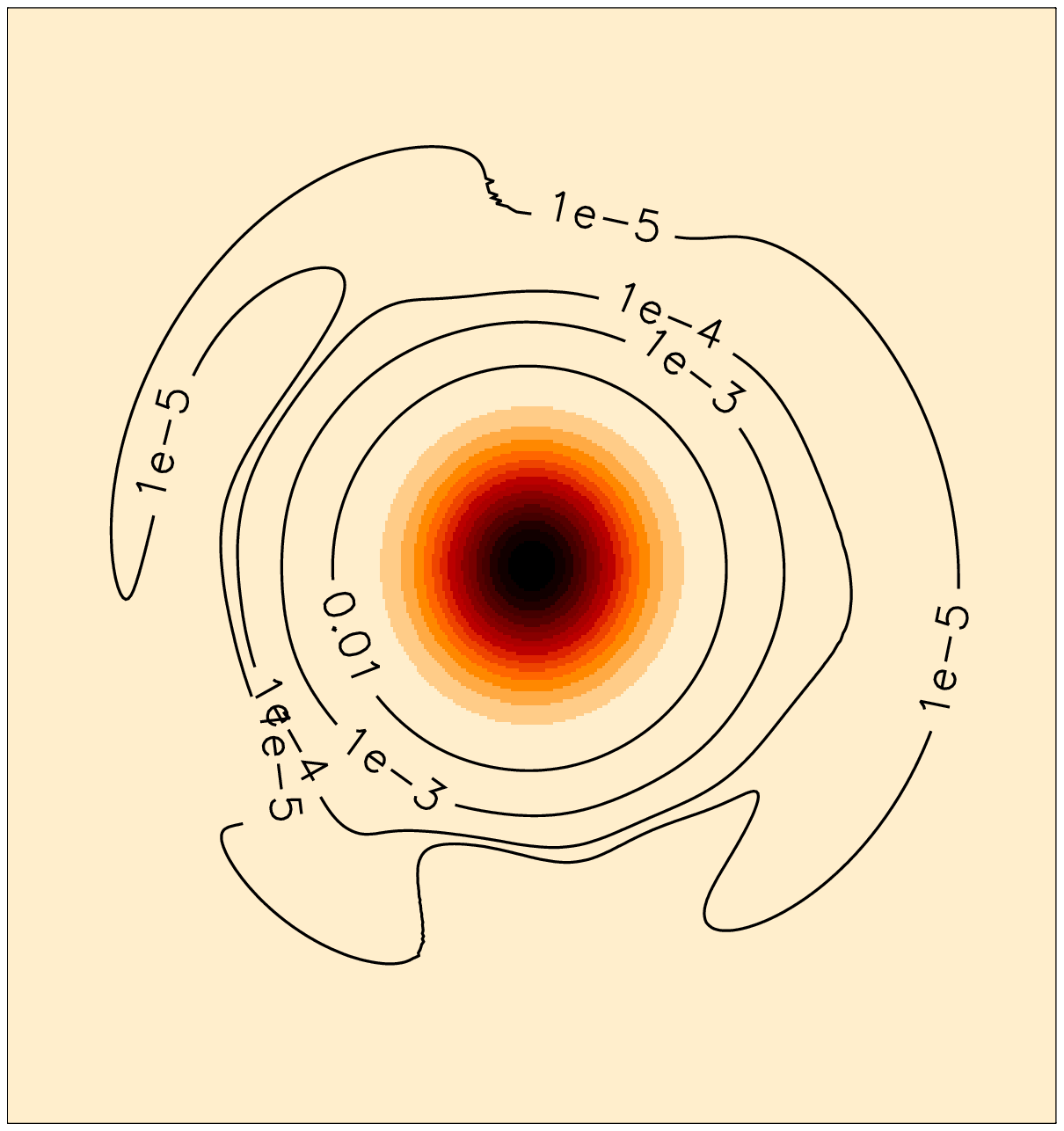}\\
\begin{tabular}{c}
$\sqrt{n_*}\mathcal{S}_{\rm eff}=10^3$\\
$(\Psi_{\rm opt} =4.3$)
\end{tabular} & 
\includegraphics[scale=0.25,bbllx=60,bblly=38,bburx=408,bbury=408]{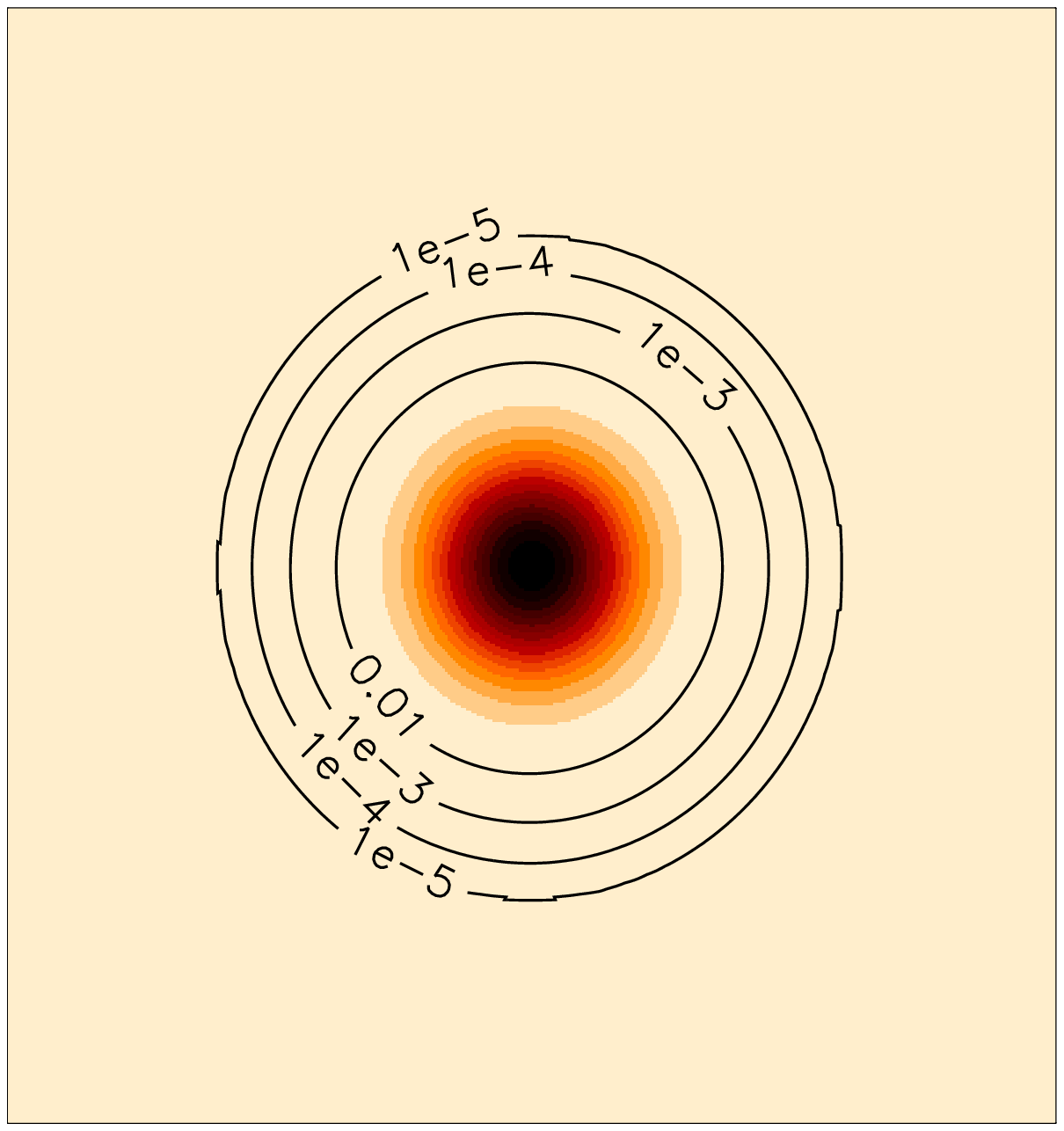} &
\includegraphics[scale=0.25,bbllx=60,bblly=38,bburx=408,bbury=408]{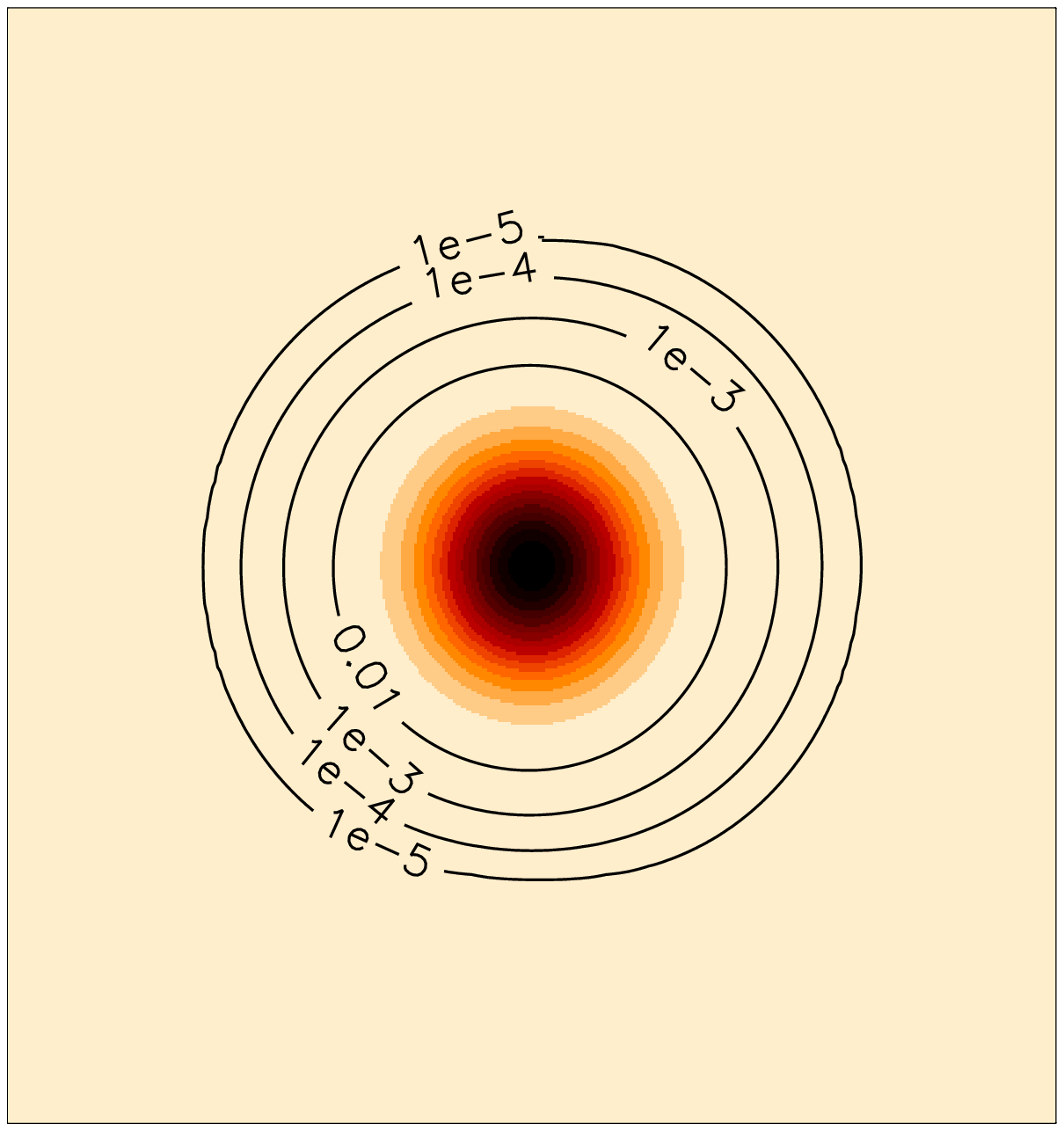} &
\includegraphics[scale=0.25,bbllx=60,bblly=38,bburx=408,bbury=408]{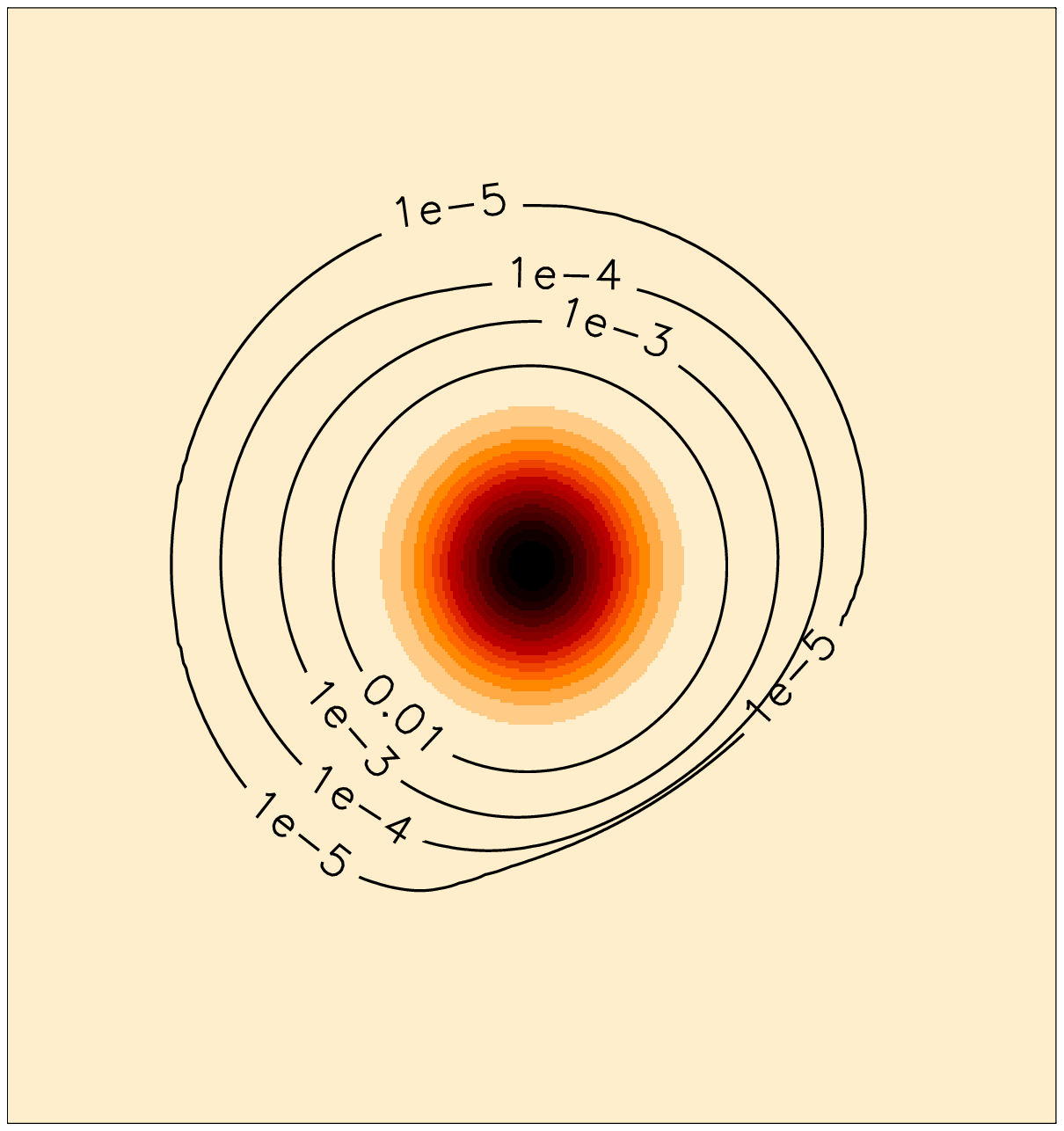} &
\includegraphics[scale=0.25,bbllx=60,bblly=38,bburx=408,bbury=408]{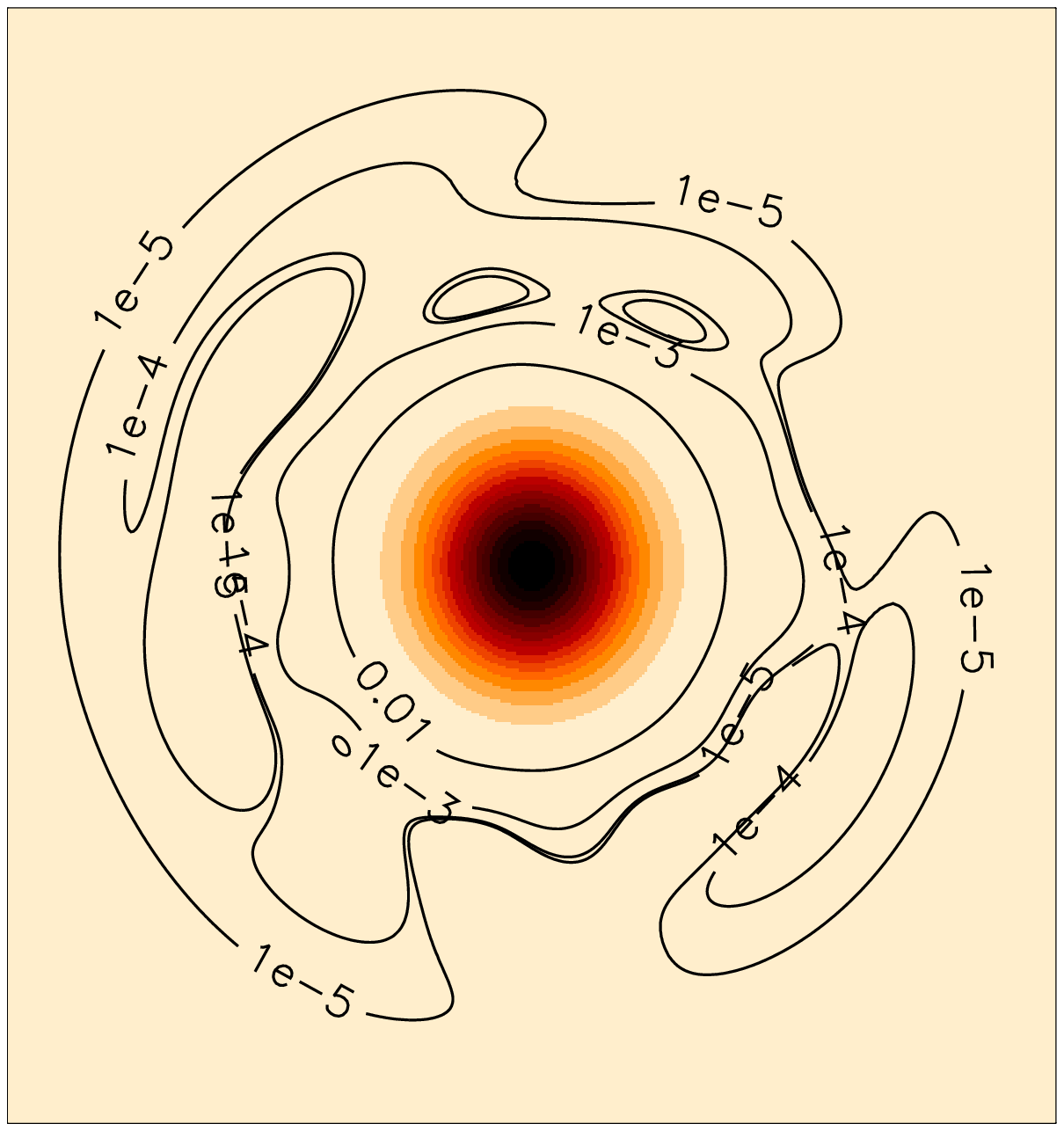}\\
\begin{tabular}{c}
$\sqrt{n_*}\mathcal{S}_{\rm eff}=100$\\
$(\Psi_{\rm opt} = 2.6$)
\end{tabular} & 
\includegraphics[scale=0.25,bbllx=60,bblly=38,bburx=408,bbury=408]{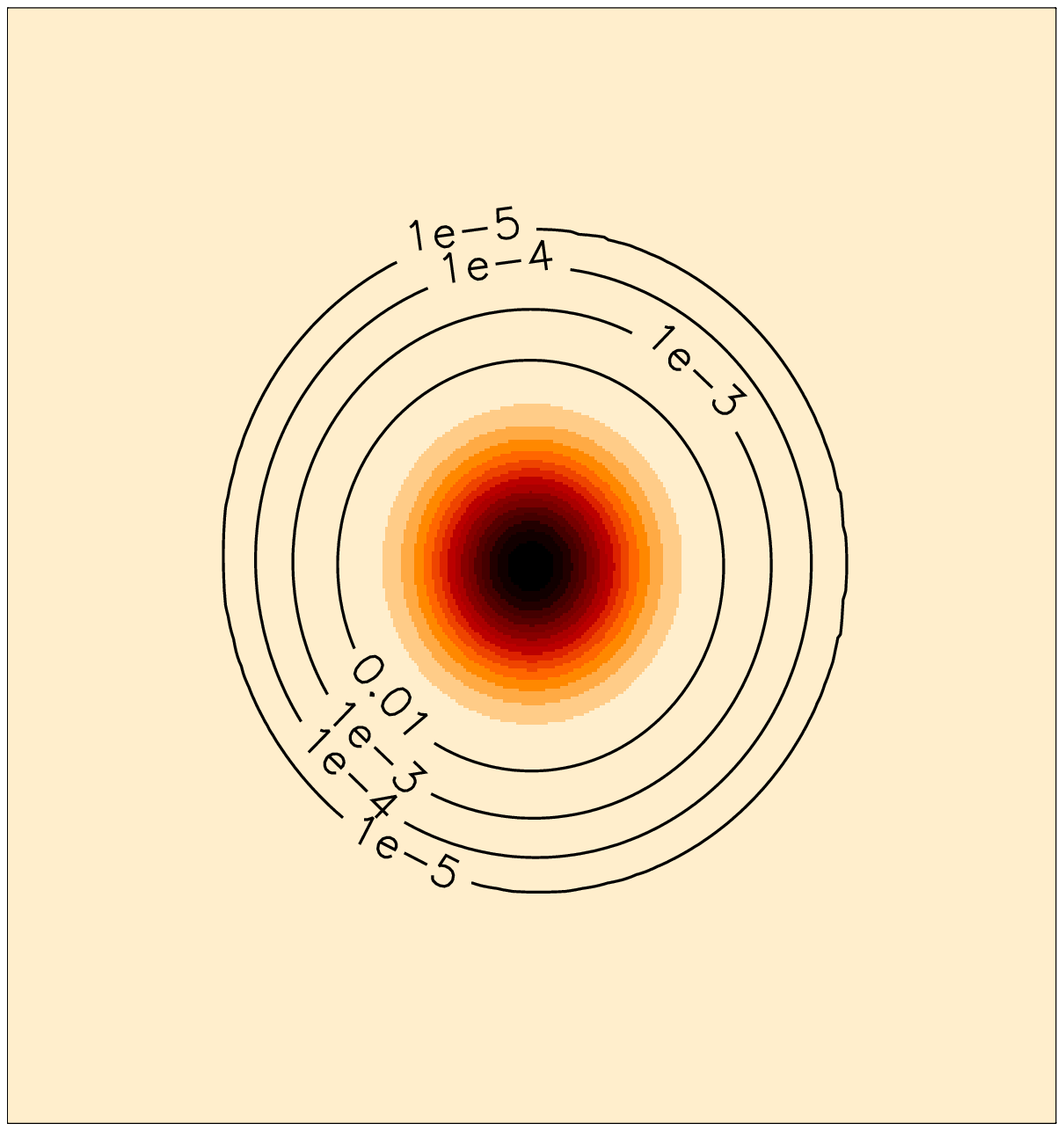} &
\includegraphics[scale=0.25,bbllx=60,bblly=38,bburx=408,bbury=408]{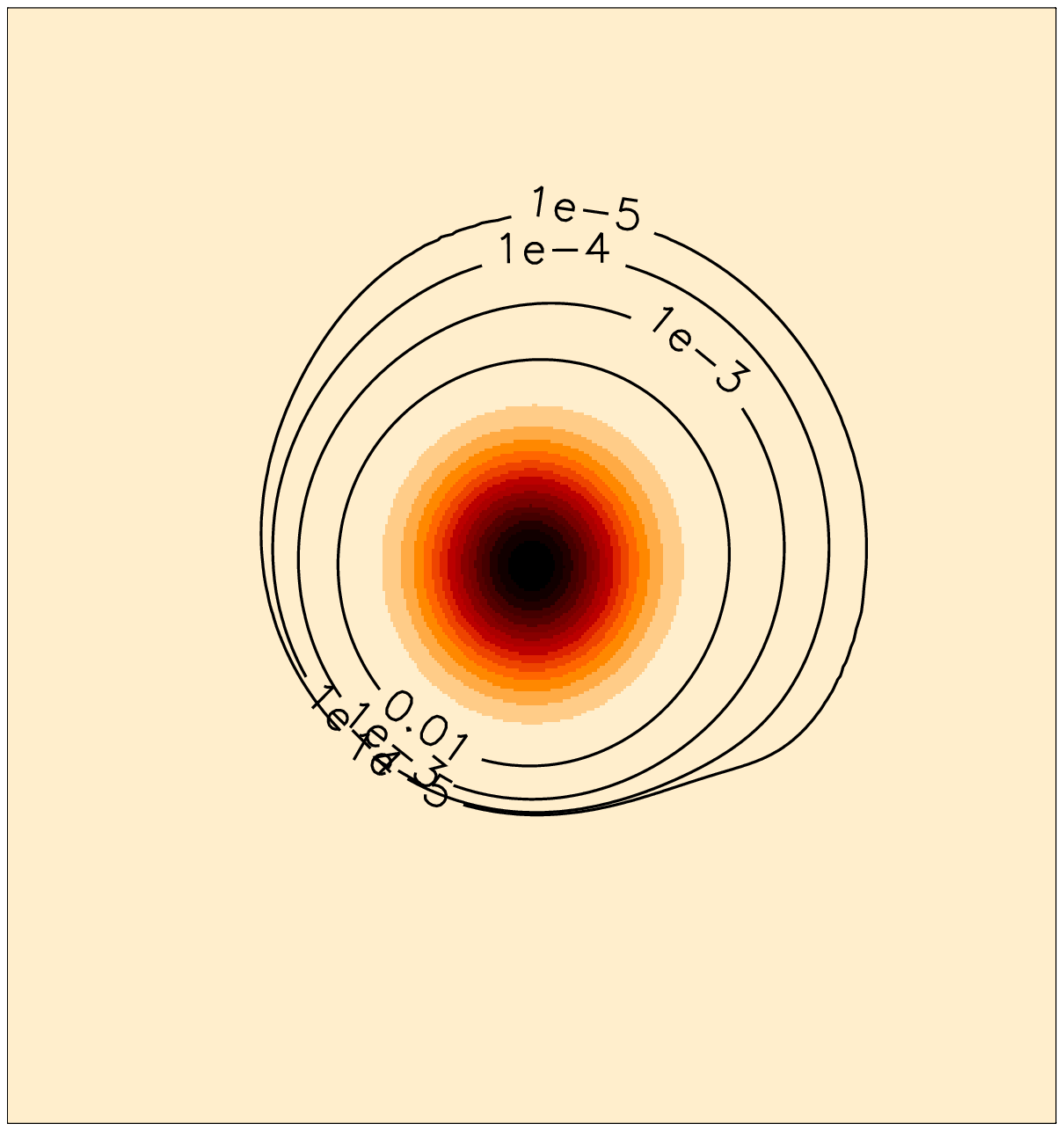} &
\includegraphics[scale=0.25,bbllx=60,bblly=38,bburx=408,bbury=408]{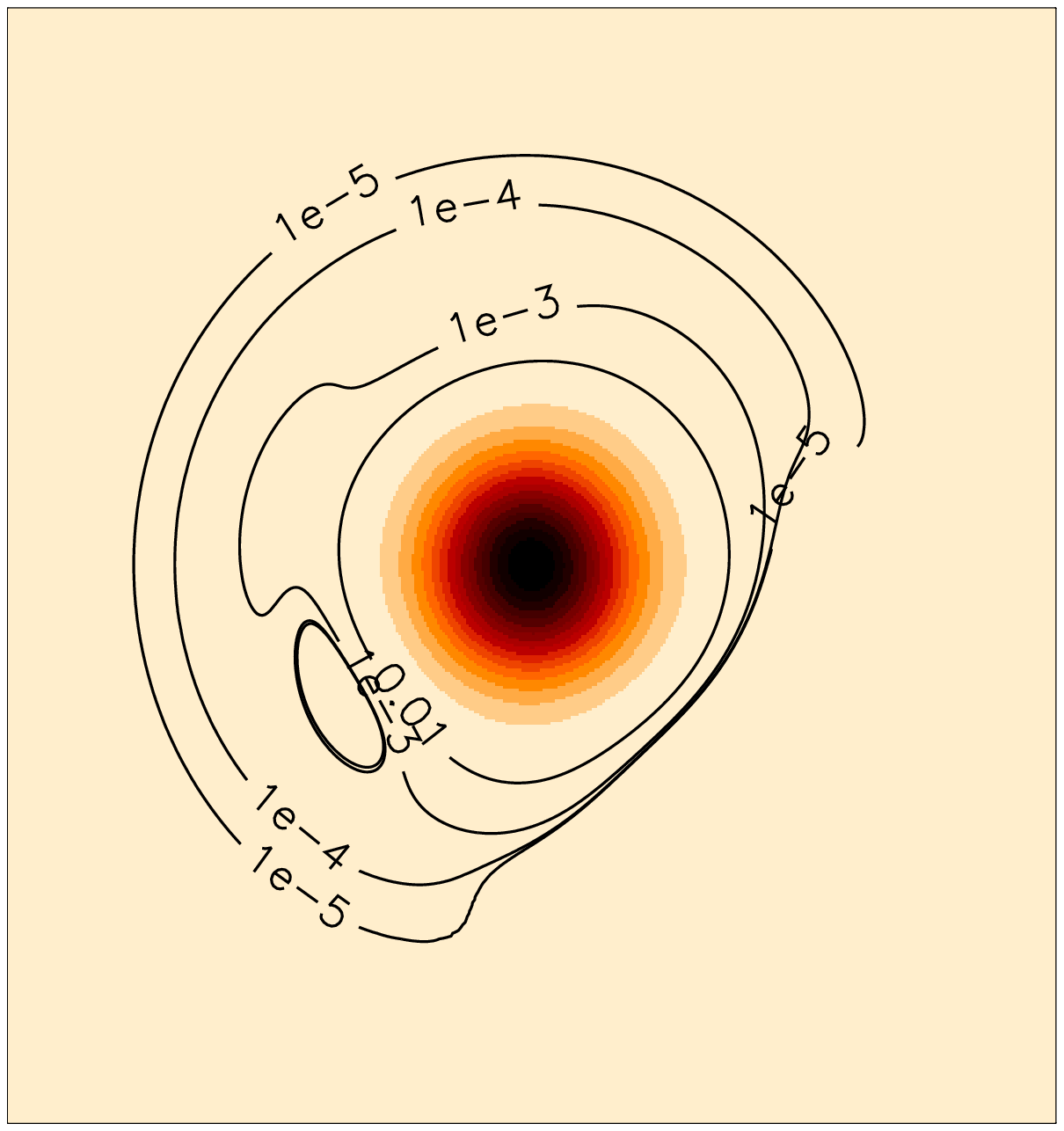} &
\includegraphics[scale=0.25,bbllx=60,bblly=38,bburx=408,bbury=408]{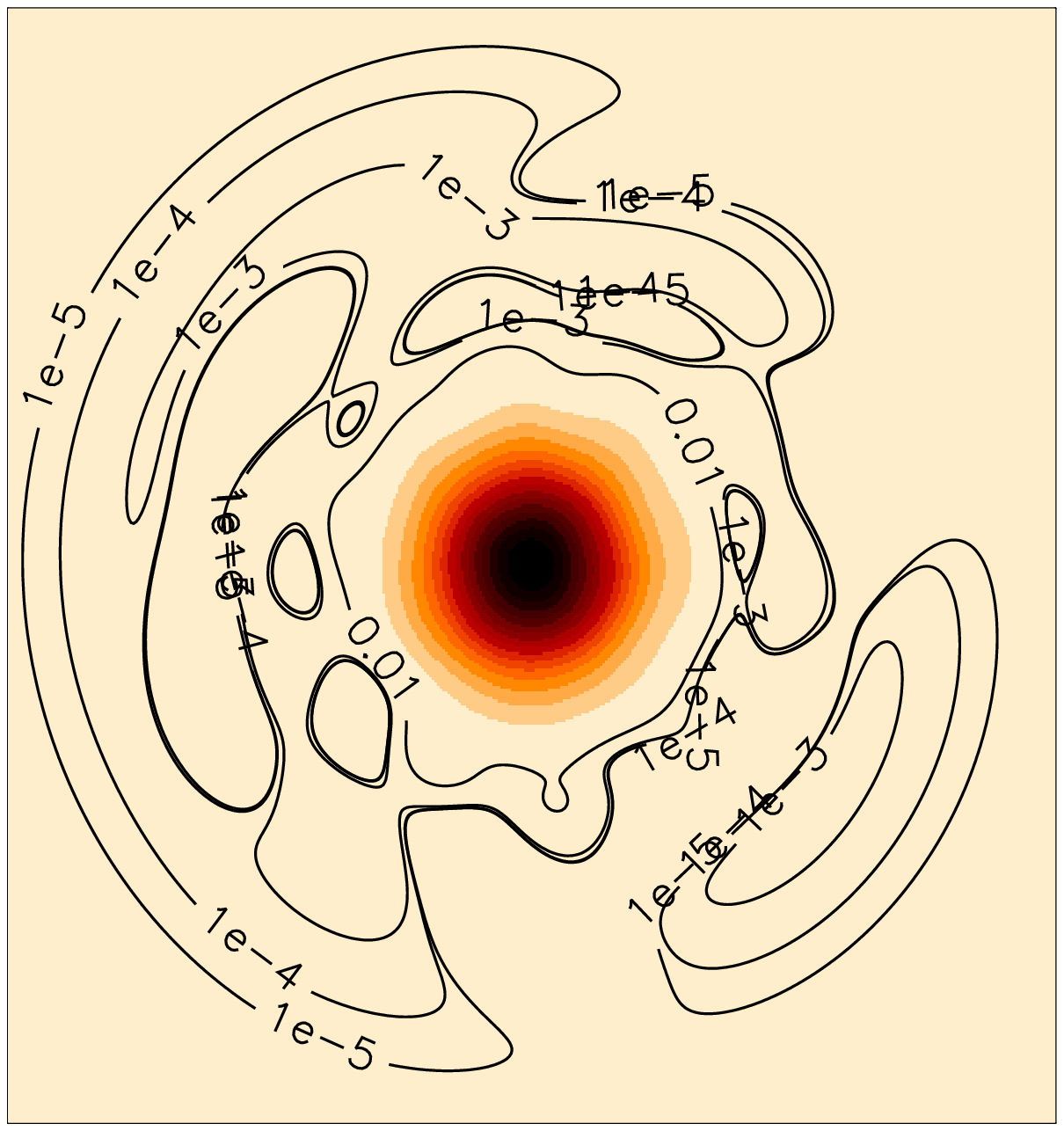}\\
\end{tabular}
\caption{
\label{fig:sparsepsf}
PSF example (top panel) adopted in this paper and best fits of it (other panels) with 4 shapelet basis sets (corresponding to $\Psi_{\rm fit}=$2.6, 4.3, 7.8 and 16.4, i.e. $n_{\rm max}=$4, 6, 10 and 20 with the diamond option) and for $\sqrt{n_*}\mathcal{S}_{\rm eff}$ equal to 100, $10^3$, $10^4$ and $\infty$ ($n_*$ is the numbers of stars used for the fit and $\mathcal{S}_{\rm eff}$ is the effective SNR of stars, see Eq. \ref{eq:snreffdef}; infinity corresponds to the ideal case of no background). 
For a given value of $\sqrt{n_*}\mathcal{S}_{\rm eff}$, all fits are performed with the same realization of the noise. 
Colors show the flux (darker colors indicate brighter regions) and show that this PSF is almost circular at the center. 
Contours show some isophotes not visible with the color scale and reveal the complex structure of the tails. 
The original (i.e. underlying) PSF in the top panel was built using a model with $\Psi=$24.8. 
The optimal value $\Psi_{\rm opt}$ of the fitted complexity (in order to minimize $\sigmasys$) is indicated under brackets for each value of $\sqrt{n_*}\mathcal{S}_{\rm eff}$. 
This figure illustrates that, for a given value of $\sqrt{n_*}\mathcal{S}_{\rm eff}$, the simpler the model (i.e. the lower $\Psi_{\rm fit}$), the poorer the description of the tails and the larger the bias. On the other hand, for a given $\Psi_{\rm fit}$, the lower the amount of information available in stars (i.e. the smaller $\sqrt{n_*}\snreff$), the noisier the description of the tails.
}
\end{figure*}

To illustrate our discussion, we study a realistic example of a PSF with complex features in the tails, 
and investigate what happens when fitting it with various shapelet basis sets as function of the SNR of the available stars. 
We also use a shapelet basis set 
(which differs from that used in the fits) for describing the underlying PSF. 
This use of shapelets for both the PSF model and the underlying PSF was chosen for three reasons:
\begin{itemize}
\item first, it allows pixelation issues to be ignored, which are beyond the scope of this paper. Indeed, the description of the underlying PSF is performed by using the continuous shapelet functions and the fits are performed at high resolution;
\item second, it considerably simplifies both the calculations and the fitting process, due to the orthogonality of shapelet functions (the average estimation of a fitted coefficient is the true value, independently of the other coefficients);
\item third, it is a simple and convenient framework for illustrating the use of sparsity as a tool in optimizing the complexity of the PSF modeling.
\end{itemize}
Our example of an underlying PSF is constructed using $n_{\rm max}=34$ (with the diamond option), as shown in Fig. \ref{fig:sparsepsf}.

In Fig. \ref{fig:sparsepsf}, we also show the 16 fits performed with the 4 shapelet basis sets (corresponding to \mbox{$n_{\rm max}=4$, 6, 10}, and 20) and $\sqrt{n_*}\snreff$ (the stellar signal-to-noise ratio, see Eq. \ref{eq:snreffdef}) equal to 100, $10^3$, $10^4$, or $\infty$ (the latter is the ideal case of no background).
To determine the overall complexity $\Psi$, which depends on the rms of galaxy ellipticities in terms of the parameter $\mathcal{E}$ (see Eqs. \ref{eq:edef} and \ref{eq:bigpsidef}), we adopt the typical value \mbox{$\mathcal{E}=0.2$}, for which \mbox{$n_{\rm max}=4$, 6, 10, 20, 34} correspond to \mbox{$\Psi=$2.6, 4.3, 7.8, 16.4, 28.4} respectively. 
In the following, we also adopt the value \mbox{$\mathcal{C}=0.066$}, that corresponds to the typical values \mbox{$P^\gamma=1.84$} and \mbox{$\left[\left<\left(\frac{\size_{\rm gal}}{\size_{\rm PSF}}\right)^4\right>\right]^{1/4}=1.5$}. 
Fig.$\,$\ref{fig:sparsepsf} illustrates that:
\begin{itemize}
\item when $\sqrt{n_*}\mathcal{S}_{\rm eff}$ is sufficiently high, complex basis sets are required to model the complex tails, i.e. the amount of bias $B$ decreases as the complexity $\Psi$ of the model increases. 
For instance, a fit with $\mathcal{S}_{\rm eff}=\infty$ and $\Psi=28.4$ would allow one to recover our PSF example exactly, with $B=0$.
\item a higher complexity requires a higher number of DoF to be fitted. Consequently, 
for a given value of $\sqrt{n_*}\mathcal{S}_{\rm eff}$, increasing the complexity of the model also increases the scatter in the estimated shape. 
Therefore, it is not always appropriate to use a complex fit model; it may be more robust to use a simplified (but more biased) fit model.
\end{itemize}


\subsection{Optimizing the complexity of the PSF model}
\label{sec:optpsfmod}
\begin{figure}
\resizebox{\hsize}{!}{
\includegraphics{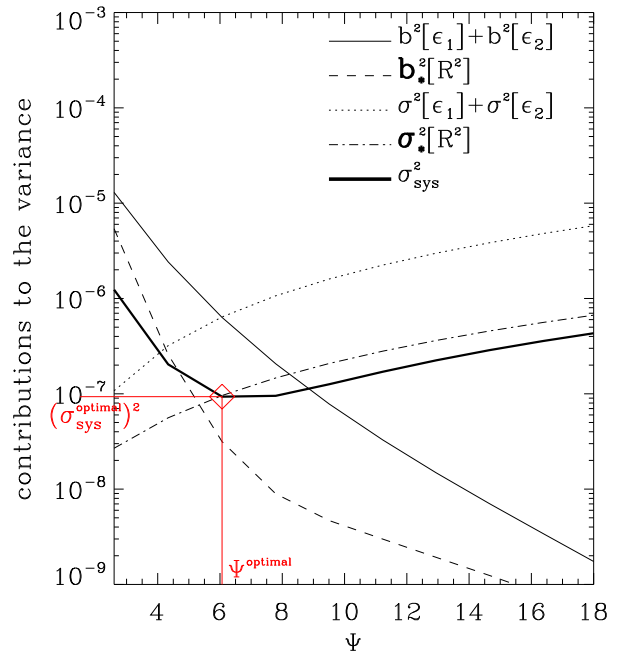}
}
\caption{
\label{fig:sigmasys.vs.psi}
Total variance $\sigma_{\rm sys}^2$ and its contributions (see Eq. \ref{eq:sigmasysdef2}) with respect to the fit model complexity $\Psi$, 
for our PSF example (see Sect.$\,$\ref{sec:psfex} and Fig.$\,$\ref{fig:sparsepsf}) fitted with 50 stars (i.e. $n_*=50$), and $\snreff=1000$. 
We note \mbox{$\bs{b_*}[\size^2]=\sqrt{\mathcal{E}}\,b[\size^2]/R^2$} and \mbox{$\bs{\sigma_*}[\size^2]=\sqrt{\mathcal{E}}\,\sigma[\size^2]/R^2$}. 
The optimal basis set is that for which $\sigma_{\rm sys}$ is minimum. 
At this point, shown by the red diamond, \mbox{$\sigma_{\rm sys}\equiv\sigma_{\rm sys}^{\rm opt}\simeq 10^{-7}$} and \mbox{$\Psi\equiv\Psi_{\rm opt}\simeq 6$}.
}
\end{figure}

The optimal PSF model is that for which $\sigma_{\rm sys}$ is minimized, varying $\Psi$. We define the optimal value $\sigma_{\rm sys}^{\rm opt}$ to be the minimal value of $\sigma_{\rm sys}$ and in the same spirit, we note that $\Psi_{\rm opt}$ is the corresponding value of $\Psi$:
\be
\label{eq:sigmasysoptdef}
\sigma_{\rm sys}^{\rm opt} \equiv \sigma_{\rm sys}\;{\rm such}\;{\rm that}\;\left.\frac{\partial (\sigma_{\rm sys}^2)}{\partial \Psi}\right|_{\Psi_{\rm opt}}=0\;.
\ee
For instance, Fig.$\,$\ref{fig:sigmasys.vs.psi} illustrates the search for the optimal shapelet basis when our PSF example (see the previous section) is estimated with 50 stars (and with $\snreff=1000$). 
The optimal model is that corresponding to $\Psi_{\rm opt}\simeq 6$ (i.e. $n_{\rm max}=8$ with the diamond option) and \mbox{$(\sigma_{\rm sys}^{\rm opt})^2\simeq 10^{-7}$}, shown by the red diamond.

\begin{figure}
\resizebox{\hsize}{!}{
\includegraphics{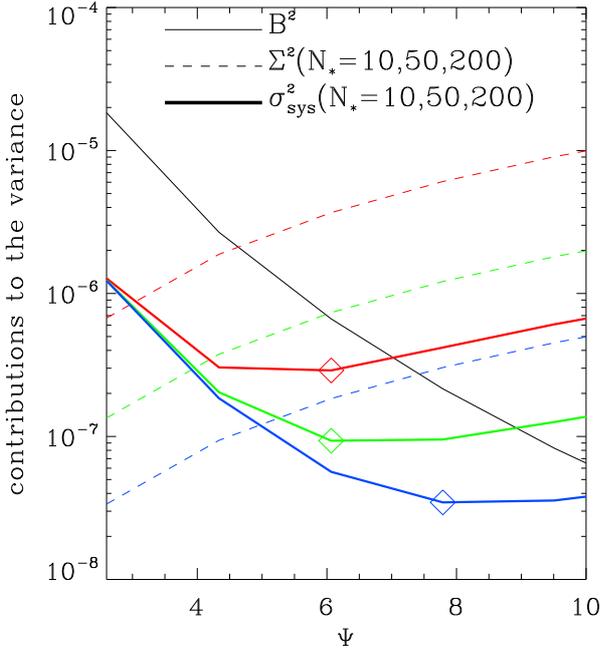}
}
\caption{
\label{fig:sigmasys.vs.psi2}
Total variance $\sigma_{\rm sys}^2$ (3 thick and solid colored curves) and its contributions to $B$ (one thin and solid black curve) and $\Sigma$ (3 dashed colored curves), as defined by Eqs. \ref{eq:tapta} to \ref{eq:sigmasysdef5}, with respect to the PSF model complexity $\Psi$, for our PSF example (presented in Sect.$\,$\ref{sec:psfex} and Fig.$\,$\ref{fig:sparsepsf}) 
fitted with 10 (red curves), 50 (green curves) or 200 (blue curves) stars (ie. $n_*=$10, 50, 200) with an effective SNR of $\snreff=1000$. 
$\Sigma$ depends on $n_*$ (see Eq. \ref{eq:totpo}), implying a different curve for each value of $n_*$, while B does not. This illustrates that 
$\sigma_{\rm sys}^{\rm opt}$ (i.e. the minimum value of $\sigma_{\rm sys}$ shown by the diamonds) increases with $n_*$.
}
\end{figure}

For a given fit model, increasing $n_*$ reduces the scatters but not the biases in the model fitting (see Eq. \ref{eq:sigmasysdef3}). Therefore, as $n_*$ increases, $\Psi_{\rm opt}$ increases and $\sigma_{\rm sys}^{\rm opt}$ decreases. 
This is illustrated in Fig. \ref{fig:sigmasys.vs.psi2}, which shows $B$, $\Sigma$, and $\sigma_{\rm sys}$ (see Eqs. \ref{eq:tapta} to \ref{eq:sigmasysdef5}) when our PSF example is estimated with $n_*=$10, 50, or 200 (still $\mathcal{S}_{\rm eff}=1000$). 

\begin{figure}
\resizebox{\hsize}{!}{
\includegraphics{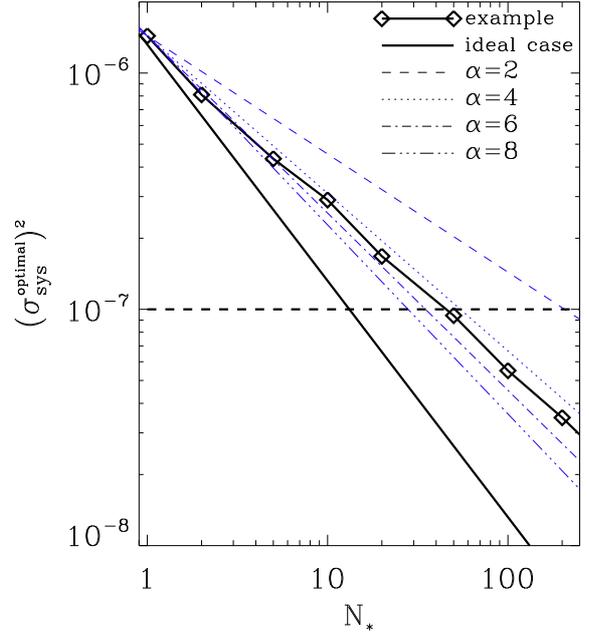}
}
\caption{
\label{fig:sigmasysoptimal.vs.nstars}
Optimal variance of the shape measurement systematics $\left(\sigma_{\rm sys}^{\rm opt}\right)^2$ 
as a function of the number of stars $n_*$ used to calibrate the PSF. The diamonds show the curve for our PSF example illustrated in all previous plots (presented on figure \ref{fig:sparsepsf}) and fitted with shapelets. The straight bold line shows the ideal case (addressed in \pun ) of a PSF model that exactly describes the PSF (i.e. with no residual). 
The horizontal line shows the values $\left(\sigma_{\rm sys}^{\rm opt}\right)^2=10^{-7}$ which is the requirement to be able to constraint $w_0$ and $w_a$ at 0.02 and 0.1 respectively \citep{2007arXiv0710.5171A}. 
The blue lines (dashed, dotted and dotted-dashed) are discussed in section \ref{sec:sparsity}. They are the curves expected when modeling the bias $B$ with a power-law function of the complexity as stated by Eq. \ref{eq:hyppowerlaw} and illustrated in Fig. \ref{fig:hyppowerlaw}.
}
\end{figure}

Fig. \ref{fig:sigmasysoptimal.vs.nstars} shows $\left(\sigma_{\rm sys}^{\rm opt}\right)^2$ as a function of $n_*$. 
The diamonds represent the curve for our PSF example illustrated in all previous plots, while the bold-straight line without any diamond shows the ideal case (addressed in \pun ) of a PSF described perfectly by the model (i.e. $B=0$). 
Thus, $\left(\sigma_{\rm sys}^{\rm opt}\right)^2$ varies with $1/n_*$ as predicted by our scaling relation presented in \pun . 
The dotted and dashed lines are discussed in Sect. \ref{sec:sparsity}.


\subsection{Example of optimal complexity in the case of a power-law function}
\label{sec:sparsity}

\begin{figure}
\resizebox{\hsize}{!}{
\includegraphics{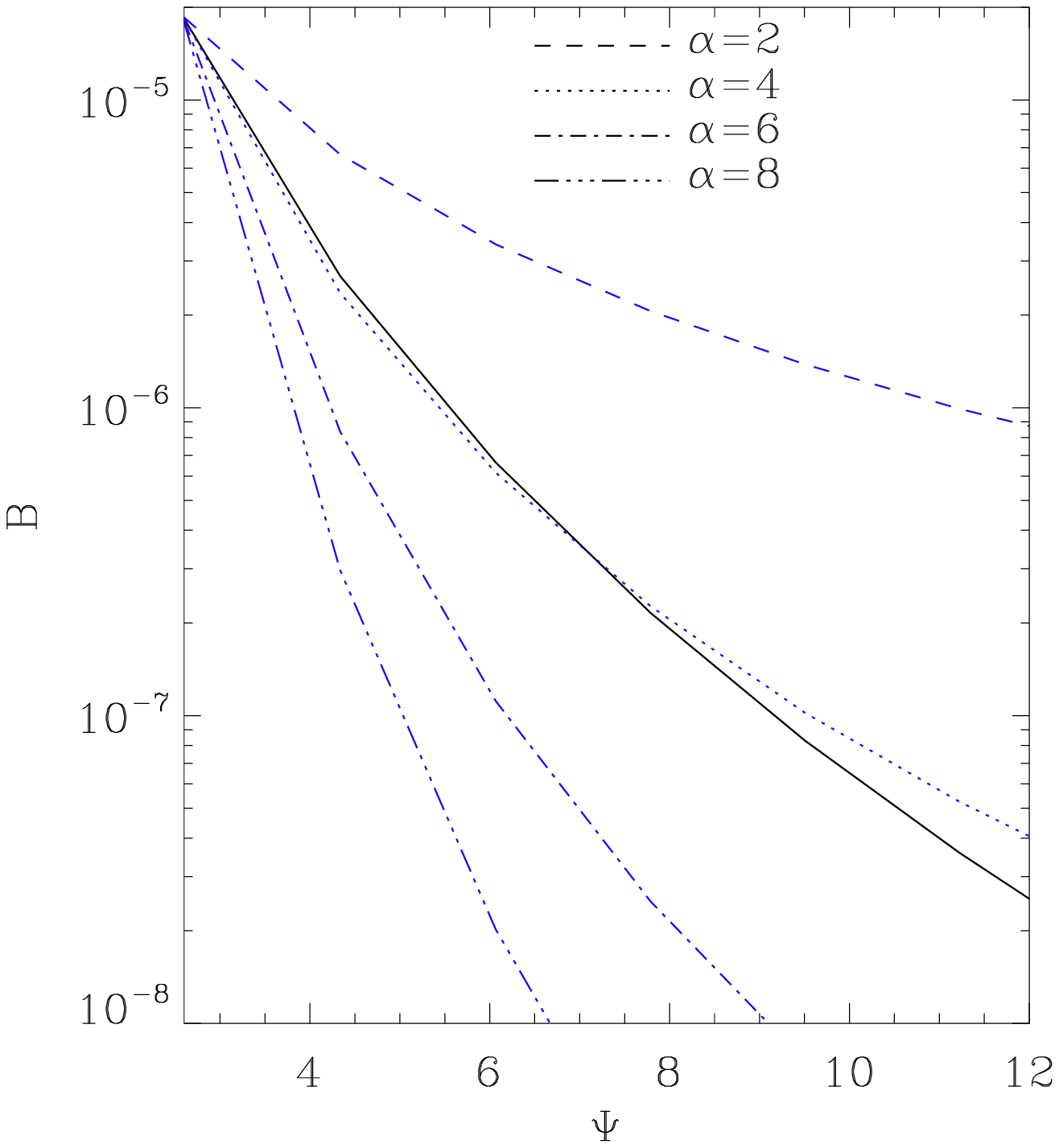}
}
\caption{
\label{fig:hyppowerlaw}
Overall bias $B$ versus $\Psi$ for our PSF example (in black) and some power-law functions $B\propto1/\Psi^\alpha$, $\alpha=$2, 4, 6, 8 (in blue) normalized to intercept at $\Psi=1.6$. 
We see that the case $\alpha=4$ fits well with the example.
}
\end{figure}

In this section, we derive the optimal complexity when the bias $B$ is a power-law function of the complexity written as $B \propto 1/\Psi^\alpha$. 
We investigate $\left(\sigma_{\rm sys}^{\rm opt}\right)^2$ as a function of $n_*$ and $\alpha$. 
We normalize the power-law function, such as:
\be
\label{eq:hyppowerlaw}
B(\Psi)\equiv B_0\,\left(\frac{\Psi_0}{\Psi}\right)^\alpha\,.
\ee
In our example of a PSF fitted with a shapelet basis set (see Sect. \ref{sec:psfex} and Fig. \ref{fig:sparsepsf}), the smallest value of $\Psi$ that we consider is 2.6 (which corresponds to $n_{\rm max}=4$ with the `diamond' configuration, see \pun ). This explains why, for this example, we choose to normalize the power-law function to this value $\Psi_0=2.6$, implying $B_0=2\times 10^{-5}$. 
This model is illustrated in Fig. \ref{fig:hyppowerlaw} for $\alpha$ equal to 2, 4, 6, and 8. 
This is superimposed on the $B$ versus $\Psi$ relation that we obtain when fitting our PSF example with different shapelet basis-sets, as decribed in Sect. \ref{sec:psfex}. We see that in this case $B$ is reasonably described by $\alpha=4$.

This representation by a power-law function is particularly convenient because $\alpha$ can be identified with the sparsity: a high value of $\alpha$ means that the PSF model is efficient in representing the underlying PSF with a small number of free parameters. Conversely, a low value of $\alpha$ means the PSF model requires a large number of parameters to describe the underlying PSF without large residuals. 
In the following, $\alpha$ is called the `sparsity parameter'. 
Together with this power-law representation (Eq.$\,$\ref{eq:hyppowerlaw}), Eqs. \ref{eq:sigmasysdef3} and \ref{eq:sigmasysoptdef} imply that:
\begin{eqnarray}
\label{eq:eq10}
\left(\sigma_{\rm sys}\left(\Psi\right)\right)^2 & = & \mathcal{C}\left[B_0\left(\frac{\Psi_0}{\Psi}\right)^\alpha + \frac{\Psi^2}{n_*\snreff^2}\right]\,,\\
\label{eq:eq12}
\Psi_{\rm opt} & = & \left[\alpha\,B_0\,\Psi_0^\alpha\,n_*\,\snreff^2/2\right]^{1/(\alpha+2)}\,,\\
\left(\sigma_{\rm sys}^{\rm opt}\right)^2 & = & \left(\sigma_{\rm sys}\left(\Psi=\Psi_{\rm opt}\right)\right)^2\nonumber\\ 
\label{eq:11}
 & = & \mathcal{C}\left[B_0^2\left(\frac{2\,\Psi_0^2}{\alpha\,n_*\mathcal{S}_{\rm eff}^2}\right)^\alpha\right]^{1/(\alpha+2)}\,.
\end{eqnarray}
Note that Eq. \ref{eq:11} expresses $\left(\sigma_{\rm sys}^{\rm opt}\right)^2$ (the minimum variance in the systematic errors in shear measurements that can be achieved, see Eqs. \ref{eq:sigmasysdef6}, \ref{eq:sigmasysdef1}, and \ref{eq:sigmasysoptdef}) in terms of a set of parameters that can be divided into 2 families:
\begin{enumerate}
\item parameters that are properties of the data set, such as $\mathcal{C}$, $B_0$, $\Psi_0$, and $\mathcal{S}_{\rm eff}$.
\item parameters that are properties of the analysis method, such as $n_*$ (i.e. the number of stars used to calibrate the PSF) and $\alpha$ (i.e. the sparsity parameter of the PSF model).
\end{enumerate}
When analysing a given data set, the parameters in the first family are kept fixed. The only parameters that can vary to achieve optimization during the analysis are those in the second family (i.e. $n_*$ and $\alpha$). 
For a given $n_*$, 
$\left(\sigma_{\rm sys}^{\rm opt}\right)^2$ is proportional to $\alpha^{-\alpha/(\alpha+2)}$.


\section{Required number of stars}
\label{sec:reqnumstars}

As discussed in the introduction, an important issue for cosmic shear surveys is to ensure that systematics are kept smaller than the statistical errors, by demanding an upper limit to $\sigma_{\rm sys}$. 
Part of the systematics have their origin in the PSF calibration, which is imperfect due to the limited number of stars available. 
In this section, we express $N_*$, the number of stars required to calibrate the PSF, in terms of the level of systematic errors $\sigma_{\rm sys}$ 
(note the capital 'N', as opposed to '$n_*$' which is the number of stars involved in the PSF calibration process: we need $n_*\ge N_*$ to ensure that systematic effects are below $\sigma_{\rm sys}$. $N_*$ is the lower limit of $n_*$).

In Sect.$\,$\ref{sec:linkp1}, we summarize the conclusions of \pun{} which apply when the underlying PSF and the PSF model have the same functional form (i.e. $B=0$) and we extend these conclusions to the general case of PSF modeling performed with any model (i.e. $B$ not necessarily equal to 0). 
In Sect.$\,$\ref{sec:reqnumstars2}, we invert Eq.$\,$\ref{eq:11} (that holds when $B$ is described by a power-law function of the complexity: $B\propto1/\Psi^\alpha$) and express $N_*$ as a function of $\alpha$ and of the minimum systematic level $\sigma_{\rm sys}^{\rm opt}$ achievable when the complexity of the PSF modeling is optimal.


\subsection{Generalised scaling relation}
\label{sec:linkp1}

In the optimistic case where the PSF calibration is the only significant source of systematic errors, a given value of $N_*$ (i.e. a given number of stars involved in the PSF calibration) implies a value of $\sigma_{\rm sys}$. This is presented in \pun{} in the form of a scaling relation 
that links $N_*$, $\sigma_{\rm sys}$, $\snreff$ (the effective signal-to-noise ratio of stars), $\left(R_{\rm gal}/R_{\rm PSF}\right)_{\rm min}$ (the ratio between the smallest galaxy size and the PSF size), and $\Psi$ (the complexity of the PSF):
\be
\label{eq:oldscalingrelation}
N_* \simeq 50
   {\left(\frac{\snreff}{500}\right)\!\!}^{-2} \!   {\left(\frac{\left(\!\size_{\rm gal}/\size_{\rm PSF}\!\right)_{\rm min}\!\!}{1.5}\right)\!\!}^{-4} \!
   {\left(\!\frac{\sigmasys^2\!}{10^{-7}}\right)\!\!\!}^{-1}\!
      {\left(\!\frac{\Psi}{3}\!\right)\!}^2 / 2\,.
\ee
The factor 2 at the end comes from the fact that $\Psi^2 \simeq 2\psi_\ellipticity^2$ (in \pun , this scaling relation is written in terms of $\psi_\ellipticity$). 
This holds for the assumption that the PSF model is able to describe the PSF without any bias (i.e. $B=0$). 
With non-zero $B$ and adopting the same simplifications and the same typical values as in \pun , Eq. \ref{eq:sigmasysdef3} leads to the more general relation: 
\be
\label{eq:newscalingrelation}
N_* \simeq 50
   {\left(\frac{\snreff}{500}\right)\!\!}^{-2} \!
   {\left(\frac{\left(\!\size_{\rm gal}/\size_{\rm PSF}\!\right)_{\rm min}\!\!}{1.5}\right)\!\!}^{-4} \!
   {\left(\!\frac{\sigmasys^2\!}{10^{-7}}\right)\!\!\!}^{-1}\!
   {\Phi}^2\,,
\ee
where
\be
\label{eq:phidef}
\Phi^2=\frac{\left(\Psi/3\right)^2}{2\left[1-\mathcal{C}\frac{B}{\sigmasys^2}\right]}\,.
\ee
Thus, taking $B$ into account in the scaling relation translates into the new factor $1/\left[1-\mathcal{C}\frac{B}{\sigmasys^2}\right]$ in Eq. \ref{eq:phidef}, which equals 1 when $B$ is zero (then the relation \ref{eq:newscalingrelation} is equivalent to the scaling relation given in \pun ) and is related to the ratio $\frac{B}{\sigmasys^2}$, which is the relative weight of biases in the error budget.


\subsection{Application to the power-law model}
\label{sec:reqnumstars2}

\begin{figure}
\resizebox{\hsize}{!}{
\includegraphics{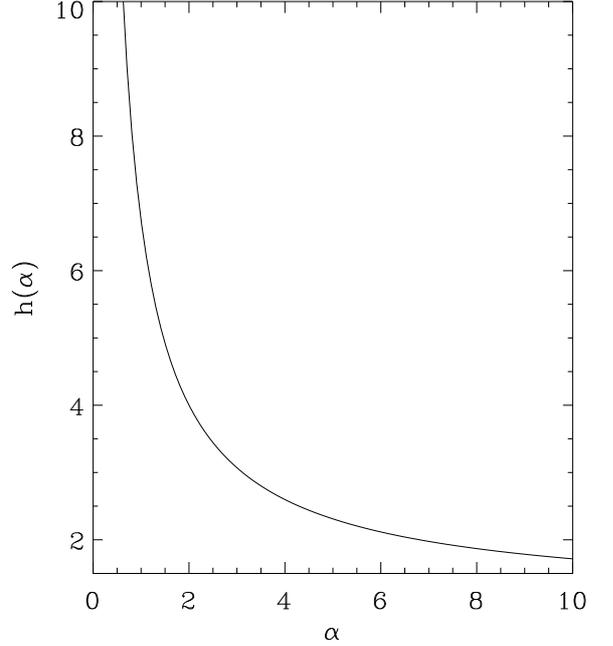}
}
\caption{
\label{fig:halpha}
$h(\alpha)$ as defined in Eq. \ref{eq:halphadef}.
}
\end{figure}

Eq. \ref{eq:11} can be inverted to provide $N_*$ (the number of stars required to calibrate the PSF) as a function of $\sigma_{\rm sys}^{\rm opt}$ (the minimum level of systematics achievable when optimizing the complexity of the PSF modeling), $\alpha$ (the sparsity parameter), $\snreff$ (the effective SNR of stars defined in Eq. \ref{eq:snreffdef}), and $\mathcal{C}$ (a dimensionless factor defined in Eq. \ref{eq:cdef}):
\be
\label{eq:nstar}
N_* = \frac{\Psi_0^2\,B_0^{2/\alpha}}{\snreff^2}\,h(\alpha) \left[\frac{\mathcal{C}}{\left(\sigma_{\rm sys}^{\rm opt}\right)^2}\right]^{(1+2/\alpha)}
\ee
with the dimensionless function:
\be
\label{eq:halphadef}
h(\alpha) = \left(\frac{\alpha}{2}\right)^{2/\alpha}\left(1+\frac{2}{\alpha}\right)^{1+2/\alpha}
\ee
shown in Fig. \ref{fig:halpha}.
With the notation and scaling of Eq. \ref{eq:newscalingrelation}, Eq. \ref{eq:nstar} is equivalent to:
\begin{eqnarray}
N_* & = & 16\left(\frac{\Psi_0}{2.6}\right)^2\left(12\,\frac{B_0}{2\times 10^{-5}}\right)^{2/\alpha}\,h(\alpha)\,\left(\frac{\snreff}{500}\right)^{-2} \times\nonumber\\
\label{eq:newnstar}
& & \left(\frac{\left(R_{\rm gal}/R_{\rm PSF}\right)_{\rm min}}{1.5}\right)^{-4(1+2/\alpha)}\left(\frac{\left(\sigma_{\rm sys}^{\rm opt}\right)^2}{10^{-7}}\right)^{-\left(1+2/\alpha\right)}\,.
\end{eqnarray}
This equation allows one to estimate the number of stars required to calibrate the PSF and thus, with respect to the stellar density, 
the minimum scale on which the PSF calibration is possible. 
On smaller scales, stars provide insufficient information for calibrating PSF variations. 
This implies that these smaller scales may be contaminated by systematics due to a poor correction of the PSF and should not be used to estimate cosmological parameters, unless the variabilities on small scales are known to be extremely small. 
As shown by \cite{2007arXiv0710.5171A} and discussed in \pun , future all-sky cosmic shear surveys will need to achieve $\left(\sigma_{\rm sys}^{\rm opt}\right)^2\le 10^{-7}$ to be able to estimate $w_0$ and $w_a$ with uncertainties of about 0.02 and 0.1, respectively. 
Fig. \ref{fig:sigmasysoptimal.vs.nstars} shows that, for our PSF example, it is possible to achieve this accuracy when calibrating the PSF with 50 stars, if $\alpha\ge 4$. 
Although this is not a general statement (this assumes that $B$ can be described by a power-law function, see Eq. \ref{eq:hyppowerlaw}, and depends on the normalization parameter $B_0$), this is a representative example of the sparsity requirement for future cosmic shear surveys. 
On the other hand, for current cosmic shear surveys of areas $\sim 50\,$deg$^2$, we have the requirement $\left(\sigma_{\rm sys}^{\rm opt}\right)^2\le 4\times 10^{-6}$ (see \cite{2007arXiv0710.5171A} and \pun ). In this case, it is possible to calibrate the PSF with a few stars when $\alpha\ge 2$. 
This sparsity requirement is reached with the current PSF correction methods \citep[for instance that based on shapelets as in][]{2008MNRAS.385..695B}, and, assuming a star density of about 1 per arcmin$^2$, this is consistent with the presence of significant $B$ modes usually found on scales smaller than a few arcmins.


\section{Conclusions}
\label{sec:conclusion}

We explore the systematics induced in cosmic shear by the PSF calibration/correction process and 
study how to optimize the PSF model to minimize the systematic errors in cosmological parameter estimations. 
In this framework, we revisit the concept of the complexity of the PSF, defined in our previous paper (\pun ), and introduce the concept of the `sparsity' of the PSF model. 
The complexity $\Psi$ characterizes the number of degrees of freedom in the model. 
A small number of degrees of freedom corresponds to a low $\Psi$ and relates to a simple PSF model, which can be fitted to the stellar observations of low signal-to-noise ratio, but is likely to be highly biased. 
On the other hand, a large number of degrees of freedom corresponds to a high $\Psi$ and relates to a complex PSF model, 
which is expected to have a low bias but requires stellar observations of high signal-to-noise ratio to avoid large statistical scatters in the fitted parameters. 
In \pun , we related the complexity $\Psi$ of the PSF model to the systematic errors in cosmological parameter estimations. In this paper, we show how the complexity can be optimized depending on the stars available, using the concept of sparsity. 
The sparsity characterizes 
the decrease of residuals between the best fit of the PSF model on the underlying PSF, when adding degrees of freedom to the model.

In the general case, we also extend  the scaling relation, proposed in \pun , between the number of stars used to calibrate the PSF and the systematic errors in the cosmological parameter estimations. 
As discussed in \pun , this relation, with the constraint of maintaining the systematics below the statistical uncertainties when estimating cosmological parameters, infers the number of stars $N_*$ required for the PSF calibration. 
$N_*$ corresponds to the minimum scale on which the PSF modeling is accurate: 
on scales smaller than this minimum, there is insufficent information in the data to calibrate the PSF variations. 
This implies that these smaller scales may be contaminated by systematics related to a poor PSF correction and should not be used when estimating cosmological parameters (unless the variabilities are known to be small, due, for instance, to the quality of the hardware).

We consider a realistic PSF example and model the amount of bias $B$ between the PSF fit and the underlying PSF by a power-law function of the 
fitted 
complexity: $B\propto 1/\Psi^\alpha$ where $\alpha$ is the sparsity parameter. 
We find that, for this PSF, current cosmic shear analyses that cover $50\,{\rm deg}^2$ or less, need $\alpha$ to be higher than 2, which is achievable by current analysis methods. 
Thus, current cosmic shear analyses do not require a rigorous optimization of the PSF model. 
On the other hand, future cosmic shear surveys that aim to measure $w_0$ and $w_a$ to an accuracy of 0.02 and 0.1, respectively, will require $\alpha\ge 4$ to calibrate the PSF with 50 stars. 
This relation between the required number of stars $N_*$ and the accuracy of the calibration depends on the underlying PSF. This explains why these values, although corresponding to realistic orders of magnitude, cannot be assumed to represent a general result. 
Two parameters drive this relation: the amount of biases $B_0$ when fitting the underlying PSF with a PSF model of low complexity ($N_*$ being proportional to $B_0^{2/\alpha}$; in our example, $B_0=2\times 10^{-5}$), and the sparsity parameter $\alpha$ of the PSF modeling during the analysis. 
It is thus possible to optimize 
cosmic shear surveys 
at two levels: 
when optimizing the observational conditions, the PSF must be as simple and stable as possible in order to make possible its description by a low complexity model (this minimizes $B_0$); 
when analysing the data, the PSF modeling must be optimized to have as high a value of the sparsity $\alpha$ as possible.

The approach suggested in this paper is a first step toward introducing the concept of sparsity to weak lensing  shape measurements. 
We do not address issues related to the pixelation. 
Moreover, although we only address the PSF calibration, this approach is also applicable to other topics such as description of galaxy shapes. 


\section*{Acknowledgments}
We thank Jean-Luc Starck and J\'erome Bobin for useful discussions and insight on sparsity. 
We also thank Sarah Bridle and Lisa Voigt for an ongoing collaboration on weak lensing shape measurements. 
SPH is supported by the P2I program, contract number 102759. AA is supported by the Swiss Institute of Technology through a Zwicky Prize.

\bibliographystyle{aa_v6.1}
\bibliography{11061}

\begin{thebibliography}{12}
\expandafter\ifx\csname natexlab\endcsname\relax\def\natexlab#1{#1}\fi

\bibitem[{{Amara} \& {Refregier}(2007{\natexlab{a}})}]{2007MNRAS.381.1018A}
{Amara}, A. \& {Refregier}, A. 2007{\natexlab{a}}, \mnras, 381, 1018

\bibitem[{{Amara} \& {Refregier}(2007{\natexlab{b}})}]{2007arXiv0710.5171A}
{Amara}, A. \& {Refregier}, A. 2007{\natexlab{b}}, ArXiv e-prints 0710.5171

\bibitem[{{Benjamin} {et~al.}(2007){Benjamin}, {Heymans}, {Semboloni}, {van
  Waerbeke}, {Hoekstra}, {Erben}, {Gladders}, {Hetterscheidt}, {Mellier}, \&
  {Yee}}]{2007MNRAS.381..702B}
{Benjamin}, J., {Heymans}, C., {Semboloni}, E., {et~al.} 2007, \mnras, 381, 702

\bibitem[{{Berg{\'e}} {et~al.}(2008){Berg{\'e}}, {Pacaud}, {R{\'e}fr{\'e}gier},
  {Massey}, {Pierre}, {Amara}, {Birkinshaw}, {Paulin-Henriksson}, {Smith}, \&
  {Willis}}]{2008MNRAS.385..695B}
{Berg{\'e}}, J., {Pacaud}, F., {R{\'e}fr{\'e}gier}, A., {et~al.} 2008, \mnras,
  385, 695

\bibitem[{{Bridle} {et~al.}(2008){Bridle}, {Shawe-Taylor}, {Amara},
  {Applegate}, {Balan}, {Bernstein}, {Berge}, {Dahle}, {Erben}, {Gill},
  {Heavens}, {Heymans}, {High}, {Hoekstra}, {Jarvis}, {Kitching}, {Kneib},
  {Kuijken}, {Lagattuta}, {Mandelbaum}, {Massey}, {Mellier}, {Moghaddam},
  {Moudden}, {Nakajima}, {Paulin-Henriksson}, {Pires}, {Rassat}, {Refregier},
  {Rhodes}, {Schrabback}, {Semboloni}, {Shmakova}, {van Waerbeke}, {Voigt}, \&
  {Wittman}}]{2008arXiv0802.1214B}
{Bridle}, S., {Shawe-Taylor}, J., {Amara}, A., {et~al.} 2008, ArXiv e-prints
  0802.1214

\bibitem[{{Fu} {et~al.}(2008){Fu}, {Semboloni}, {Hoekstra}, {Kilbinger}, {van
  Waerbeke}, {Tereno}, {Mellier}, {Heymans}, {Coupon}, {Benabed}, {Benjamin},
  {Bertin}, {Dor{\'e}}, {Hudson}, {Ilbert}, {Maoli}, {Marmo}, {McCracken}, \&
  {M{\'e}nard}}]{2008A&A...479....9F}
{Fu}, L., {Semboloni}, E., {Hoekstra}, H., {et~al.} 2008, \aap, 479, 9

\bibitem[{{Heymans} {et~al.}(2006){Heymans}, {VanWaerbeke}, {Bacon}, {Berge},
  {Bernstein}, {Bertin}, \& {Bridle}}]{2006MNRAS.368.1323H}
{Heymans}, C., {VanWaerbeke}, L., {Bacon}, D., {et~al.} 2006, \mnras, 139, 313

\bibitem[{{Massey} {et~al.}(2007){Massey}, {Heymans}, {Berg{\'e}}, {Bernstein},
  {Bridle}, {Clowe}, {Dahle}, {Ellis}, {Erben}, {Hetterscheidt}, {High},
  {Hirata}, {Hoekstra}, {Hudelot}, {Jarvis}, {Johnston}, {Kuijken},
  {Margoniner}, {Mandelbaum}, {Mellier}, {Nakajima}, {Paulin-Henriksson},
  {Peeples}, {Roat}, {Refregier}, {Rhodes}, {Schrabback}, {Schirmer}, {Seljak},
  {Semboloni}, \& {van Waerbeke}}]{2007MNRAS.376...13M}
{Massey}, R., {Heymans}, C., {Berg{\'e}}, J., {et~al.} 2007, \mnras, 376, 13

\bibitem[{{Massey} \& {Refregier}(2005)}]{2005MNRAS.363..197M}
{Massey}, R. \& {Refregier}, A. 2005, \mnras, 363, 197

\bibitem[{{Paulin-Henriksson} {et~al.}(2008){Paulin-Henriksson}, {Amara},
  {Voigt}, {Refregier}, \& {Bridle}}]{2008A&A...484...67P}
{Paulin-Henriksson}, S., {Amara}, A., {Voigt}, L., {Refregier}, A., \&
  {Bridle}, S.~L. 2008, \aap, 484, 67

\bibitem[{{Rhodes} {et~al.}(2008){Rhodes}, {}, {}, {}, {}, {}, {}, \&
  {}}]{spacestep}
{Rhodes}, J., {}, {}, {et~al.} 2008, in prep.

\bibitem[{{Rhodes} {et~al.}(2000){Rhodes}, {Refregier}, \&
  {Groth}}]{2000ApJ...536...79R}
{Rhodes}, J., {Refregier}, A., \& {Groth}, E.~J. 2000, \apj, 536, 79

\end{thebibliography}

\end{document}